\documentclass[twocolumn]{aastex63}
\usepackage{CJK}
\usepackage{amsmath}
\usepackage{verbatim}

\usepackage{hyperref}
\usepackage{multirow}

\definecolor{blue}{RGB}{50, 80, 255}
\definecolor{red}{RGB}{255, 50, 50}

\shorttitle{Gas Disk Model}
\shortauthors{Xu et al.}

\begin{document}
\begin{CJK}{UTF8}{gbsn}
\title{Modeling Circumstellar Gas Emission around a White Dwarf Using Cloudy}

\correspondingauthor{Siyi Xu} 
\email{siyi.xu@noirlab.edu}

\author[0000-0002-8808-4282]{Siyi Xu (许\CJKfamily{bsmi}偲\CJKfamily{gbsn}艺)} 
\affil{Gemini Observatory/NSF's NOIRLab, 670 N. A'ohoku Place, Hilo, Hawaii, 96720, USA}

\author[0000-0002-4037-3114]{Sherry Yeh} 
\affil{W. M. Keck Observatory, Waimea, HI, USA}

\author[0000-0002-3553-9474]{Laura. K. Rogers}
\affil{Institute of Astronomy, University of Cambridge, Madingley Road, Cambridge CB3 0HA, UK}

\author[0000-0002-0141-7946]{Amy Steele}
\affil{Yerkes Observatory, 373 W. Geneva St., Williams Bay, WI, USA}

\author[0000-0003-2852-268X]{Erik Dennihy}
\affil{Rubin Observatory Project Office, 950 N. Cherry Ave., Tucson, AZ 85719, USA}

\author[0000-0003-0053-3854]{Alexandra E. Doyle}
\affil{Earth, Planetary, and Space Sciences, University of California, Los Angeles, Los Angeles, CA 90095, USA}

\author{P. Dufour}
\affil{D\'epartement de Physique, Universit\'e de Montr\'eal, C.P. 6128, Succ. Centre-Ville, Montr\'eal, Qu\'ebec H3C 3J7, Canada}

\author[0000-0001-5854-675X]{Beth L. Klein}
\affil{Department of Physics and Astronomy, University of California, Los Angeles, CA 90095-1562, USA}

\author[0000-0003-1543-5405]{Christopher J. Manser}
\affil{Astrophysics Group, Department of Physics, Imperial College London, Prince Consort Rd, London, SW7 2AZ, UK}
\affil{Department of Physics, University of Warwick, Coventry CV4 7AL, UK}

\author[0000-0001-9834-7579]{Carl Melis}
\affil{Center for Astrophysics and Space Sciences, University of California, San Diego, CA 92093-0424, USA}

\author[0000-0002-1517-6792]{Tinggui Wang}
\affil{CAS Key Laboratory for Researches in Galaxies and Cosmology, University of Sciences and Technology of China, Hefei, Anhui 230026, Peopleʼs Republic of
China}
\affil{School of Astronomy and Space Sciences, University of Science and Technology of China, Hefei, 230026, Peopleʼs Republic of China}

\author[0000-0001-6654-7859]{Alycia J. Weinberger}
\affil{Earth and Planets Laboratory, Carnegie Institution for Science, 5241 Broad Branch Rd NW, Washington, DC 20015, USA}

\begin{abstract}

The chemical composition of an extrasolar planet is fundamental to its formation, evolution and habitability. In this study, we explore a new way to measure the chemical composition of the building blocks of extrasolar planets, by measuring the gas composition of the disrupted planetesimals around white dwarf stars. {\color{black} As a first attempt, we used the photo-ionization code \textsc{Cloudy} to model the circumstellar gas emission around a white dwarf Gaia~J0611$-$6931 under some simplified assumptions. We found most of the emission lines are saturated and the line ratios approaching the ratios of thermal emission; therefore only lower limits to the number density can be derived. Silicon is the best constrained element in the circumstellar gas and we {\color{black} derived} a lower limit of 10$^{10.3}$ cm$^{-3}$. In addition, {\color{black} we placed a lower limit on the total amount of gas to be} 1.8 $\times$ 10$^{19}$~g. Further study is needed to better constrain the parameters of the gas disk and connect it to other white dwarfs with circumstellar gas absorption.
} 
\end{abstract}

\keywords{Circumstellar gas, Chemical abundances, Debris Disks, Planetesimals, Extrasolar Rocky Planets, White Dwarf Stars}

\section{Introduction}

The chemical composition of an exoplanet is an important property, but it is generally hard to constrain because different compositions can have very similar mass-radius relationships \citep[e.g.][]{Dorn2015}. One way to get around this limitation is by observing so called “polluted” white dwarfs, which are actively accreting from disrupted planetesimals. As a result, spectroscopic observations of the atmospheres of these white dwarfs can be used to measure individual elements, enabling the determination of the chemical compositions of extrasolar planets \citep[e.g.][]{JuraYoung2014}. To zeroth order, the planetary compositions measured from polluted white dwarfs resemble that of bulk Earth: O, Mg, Si, and Fe are the dominant elements in their familiar proportions \citep[e.g.][]{Klein2011, Xu2014, Swan2019, Trierweiler2023}. Objects with a large amount of water and volatile elements have been detected as well \citep[e.g.][]{Xu2017,Hoskin2020}. However, this kind of analysis is subject to our understanding of white dwarf physics and many areas are still under active research, e.g. radiative levitation, {\color{black} thermohaline instability}, convective overshooting, and 3D effects \citep{Koester2014a,BauerBildsten2019,Cunningham2019,Cukanovaite2021}. In addition, the measured abundance ratios can vary by up to an order of magnitude depending on the assumed accretion stage \citep{Koester2009a}.

In this study, we explore a new way to {\color{black} constrain} the chemical compositions of extrasolar planetesimals by modeling the gas debris around white dwarf stars. There are 21 white dwarfs with a debris disk that has both dust and gas emission \citep{Melis2020,Dennihy2020b,Gentile-Fusillo2021a}. Most systems display double-peaked asymmetric emission lines, which are hallmarks of a rotating eccentric disk \citep{HorneMarsh1986,Gaensicke2006,Manser2021}. They are fresh extrasolar planetary material in a transient stage, right after tidal disruption and before being completely accreted onto the white dwarf. The occurrence rate of these gas disks is estimated to be 0.06\%, likely because the gaseous components are a rare subset of white dwarf debris disks \citep{Manser2020}. In comparison, the occurrence rate of infrared excess from a debris disk around a white dwarf is about 2--4\% \citep{Barber2016,Wilson2019}. The gas disks can have complex morphologies, as shown by the Doppler tomography observations \citep[e.g.][]{Manser2016}. In addition, the morphology and strength of the emission lines can be variable \citep{Wilson2014}. For some systems, the variability is periodic and it can be explained as a product of precession under the general relativity and gas pressure forces \citep{MirandaRafikov2018,Goksu2023}. In the most extreme case of WD~J2100+2122, the emission lines appeared and disappeared within a few months \citep{Dennihy2020b}.

 Early modeling done by \citet{Hartmann2011,Hartmann2016} assumed that the gaseous material is heated by viscous accretion, which overestimates the mass accretion rates and predicts many emission lines that are not observed. The currently favored heating mechanism is photo-ionization, where the central white dwarf is the main energy source \citep{Melis2010}. This concept is demonstrated in the exciting discovery of a sulfur-rich gas disk around a white dwarf, likely from an evaporating giant planet \citep{Gaensicke2019}. The photo-ionization code \textsc{Cloudy} has also been used to model the circumstellar gas absorption around WD~1124$-$293 \citep{Steele2021}. 
 
 All the rocky gas disk hosts also show infrared excess from a dust disk. The geometrically thin and optically thick dust disk model proposed by \citet{Jura2003} is widely used. However, such a model fails to explain bright disks \citep[e.g.][]{Jura2007b} and variable disks \citep[e.g.][]{XuJura2014,Swan2020}, with the disks that host gaseous emission being among the brightest and most variable \citep{Dennihy2017,Swan2020}. At least some optically thin dust must be present in the system \citep{Ballering2022}. 

The emission line systems provide a new window to study chemical compositions of extrasolar planets. Here, we focus on the gas disk around the white dwarf \object{Gaia~J061131.70$-$693102.15} (referred to as Gaia~J0611$-$6931 for the remainder of the paper). The parameters of the white dwarf are listed in Table~\ref{tab:properties}. Gaia~J0611$-$6931 was first identified as a candidate white dwarf from the Data Release 2 of the {\it Gaia} mission \citep{Jimenez-Esteban2018, Gentile-Fusillo2019}. Follow-up spectroscopic and photometric observations show that it is indeed a white dwarf with many interesting properties. Gaia~J0611$-$6931 has a strong infrared excess with a fractional luminosity of about 5\%, much higher than the typical white dwarf debris disks \citep{Dennihy2020b}. It has a heavily polluted atmosphere and photospheric absorption from {\color{black} 10 different elements} has been detected \citep{Rogers2024}. In addition, it displays numerous emission lines in the optical and infrared spectra, i.e. O, Na, Mg, Si, and Ca, making it one of the white dwarfs showing the largest variety of gas emission species \citep{Dennihy2020b, Melis2020, Owens2023}. More importantly, the gas line profiles are symmetric and stable, making Gaia~J0611$-$6931 an ideal target for modeling.

\begin{deluxetable}{lcccccccccc}
\tablecaption{ \label{tab:properties} Properties of Gaia~J0611$-$6931 from \citet{Rogers2024}.}
\tablehead{
\colhead{Parameter} &  \colhead{Value } }
\startdata
Coordinate &  06:11:31.70 -69:31:02.15\\
Gaia ID & 5279484614703730944 \\
Spectral Type & DAZ  \\
G (mag) & 16.8\\
Temperature (K) & 17749\\
Surface Gravity (cm s$^\mathrm{-2}$) & 8.14\\
Mass (M$_\mathrm{\odot}$)& 0.702 \\
Radius (R$_\mathrm{\odot}$) & 0.0118\\
Luminosity (L$_\mathrm{\odot}$) & 0.012\\
{\color{black} Distance (pc)} & {\color{black} 143} \\
  \enddata
\end{deluxetable}

This paper is organized as follows: A summary of the observations and the characteristics of the emission lines around Gaia~J0611$-$6931 is presented in Section~\ref{sec: observations}. The overall modeling method, including \textsc{Cloudy} modeling and line profile modeling, is given in Section~\ref{sec: method}, and its application to Gaia~J0611$-$6931 is in Section~\ref{sec: result}. Discussion and conclusions are presented in Section~\ref{sec: discussion} and \ref{sec: conclusions}, respectively.

\section{Observations and Data Analysis \label{sec: observations}}

\subsection{Observations}

\begin{figure*}[tbh]
    \centering
    \includegraphics[width=0.99\textwidth]{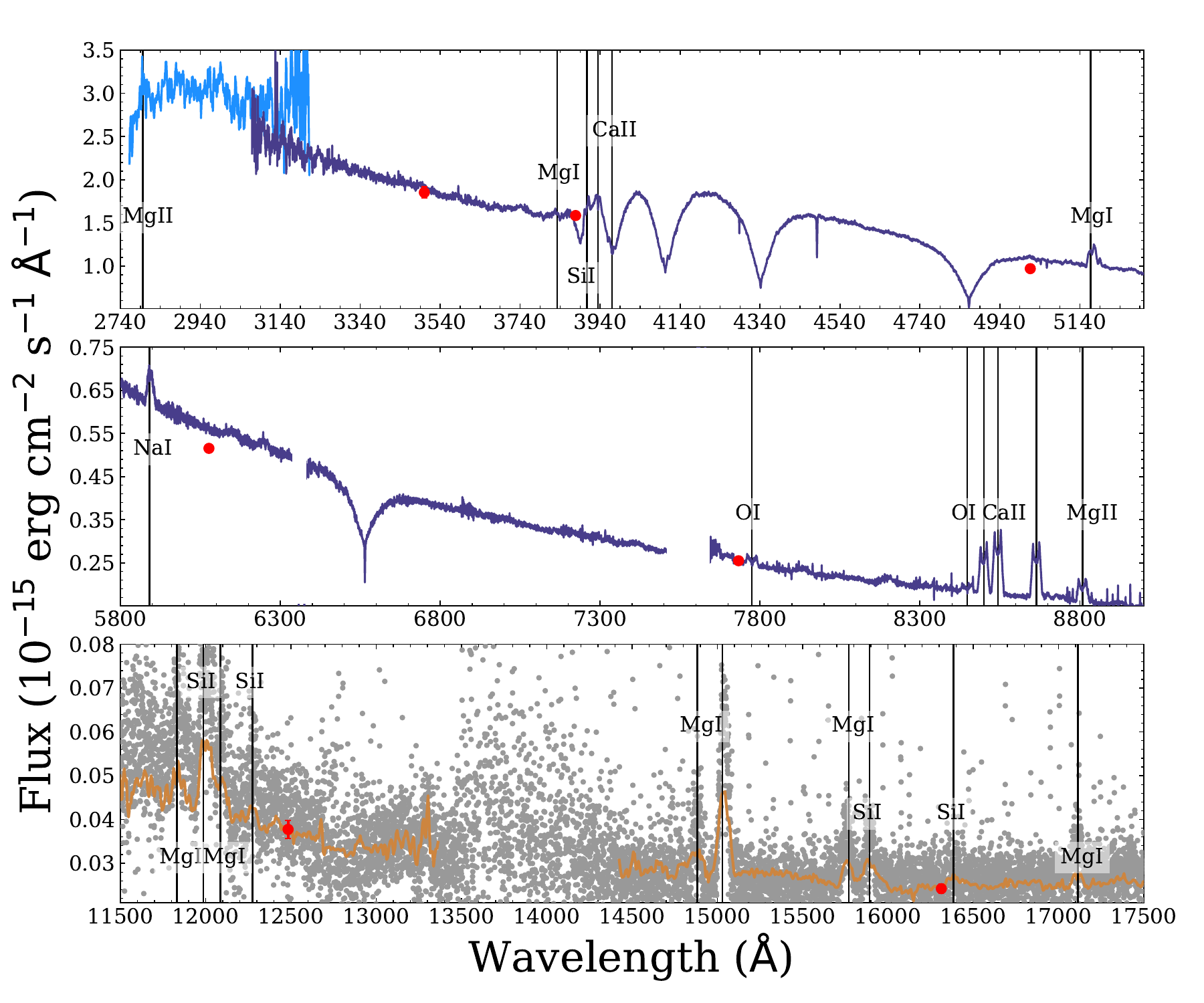}
    \caption{The complete spectra of Gaia~J0611$-$6931 with flux calibration {\color{black} anchored to} the SkyMapper photometry (red dots). The light blue, dark blue, and orange lines represent data from the HST/COS, the VLT/X-SHOOTER (UVB and VIS arms), and the Gemini/Flamingos-2, respectively. The grey dots are data from the X-SHOOTER NIR arm and are only used for facilitating line identifications. The black lines mark emission features listed in Table~\ref{tab:lines}.
    \label{fig: complete_spectra}
    }
\label{fig:spectra_all}
\end{figure*}

Gaia~J0611$-$6931 has been observed extensively using {\color{black}multiple} optical, infrared and ultraviolet (UV) spectrographs \citep{Dennihy2020b, Melis2020, Rogers2024, Owens2023}. In this paper, we focus on data from the X-SHOOTER on the Very Large Telescope (program IDs: 104.C-0107, 105.202N, and 106.2130), the Cosmic Origins Spectrograph (COS) on the Hubble Space Telescope (\citealt{Rogers2024}, ID: 16204)\footnote{\color{black} The HST/COS data presented in this paper were obtained from the Mikulski Archive for Space Telescopes (MAST) at the Space Telescope Science Institute. The specific observations analyzed can be accessed via \dataset[DOI: 10.17909/zssq-wr33]{https://doi.org/10.17909/zssq-wr33}.}, and the Flamingos-2 at the Gemini Observatory (GS-2021B-Q-244, \citealt{Owens2023}). The X-SHOOTER data were obtained on three different dates (i.e., October 15 2019, December 3, 2020, and August 22, 2021 UT), and no significant changes were identified in the shape or the strength of the emission lines. Therefore, all the spectra were combined to have the highest signal to noise ratio (S/N). Stare mode was used for the first two epochs, while nod on slit was used for the last epoch, so the spectra from the NIR arm can be extracted. We used the standard ESO data reduction pipeline esoreflex \citep{Freudling2013} and molecfit \citep{Kausch2015,Smette2015}. The complete {\color{black} spectrum} of Gaia~J0611$-$6931 is shown in Figure~\ref{fig:spectra_all}. The flux calibration is performed using photometry from SkyMapper \citep{Onken2019}. The wavelength is presented in vacuum throughout the paper.

In the infrared {\color{black} beyond 1 micron}, the X-SHOOTER data have higher spectral resolution than Flamingos-2, which is useful for checking the line identifications presented in \citet{Owens2023}. However, the individual exposure time is 300~sec for the NIR arm, which may be too long for proper sky subtraction. That might explain the apparent difference in the line strength between X-SHOOTER and Flamingos-2 data. 

As an additional check, we obtained a medium resolution (R$\approx$25,000) optical spectrum of Gaia~J0611$-$6931 with the Magellan Inamori Kyocera Echelle (MIKE) spectrograph \citep{Bernstein2003} on the 6.5-m Magellan Clay Telescope on August 27, 2021 (UT). The observation and data reduction were performed {\color{black} using \textsc{CarPy} \citet{Kelson2000,Kelson2003}}. The MIKE data provides a wavelength coverage of 3500---5060~{\AA} for the blue arm and 5000--9400~{\AA} for the red arm. The total exposure time is 3111~s and the S/N was 12 at 4000~{\AA} and 14 at 6370~{\AA}. We did not notice any difference between the X-SHOOTER and MIKE spectra.

The spectral classification of Gaia~J0611$-$6931 is DAZ, which means hydrogen lines are the strongest features in the optical spectrum, as can be seen in Figure~\ref{fig: complete_spectra}. There are also narrow absorption features from elements heavier than helium in the spectra (giving ``Z" in the spectral classification) and the abundance analysis is presented in \citet{Rogers2024}. In this paper, we focus on the emission features from the circumstellar gas around Gaia~J0611$-$6931.

\subsection{Emission Line Analysis \label{sec: line analysis}}

\begin{figure}[tbh]
    \centering
    \includegraphics[width=0.48\textwidth]{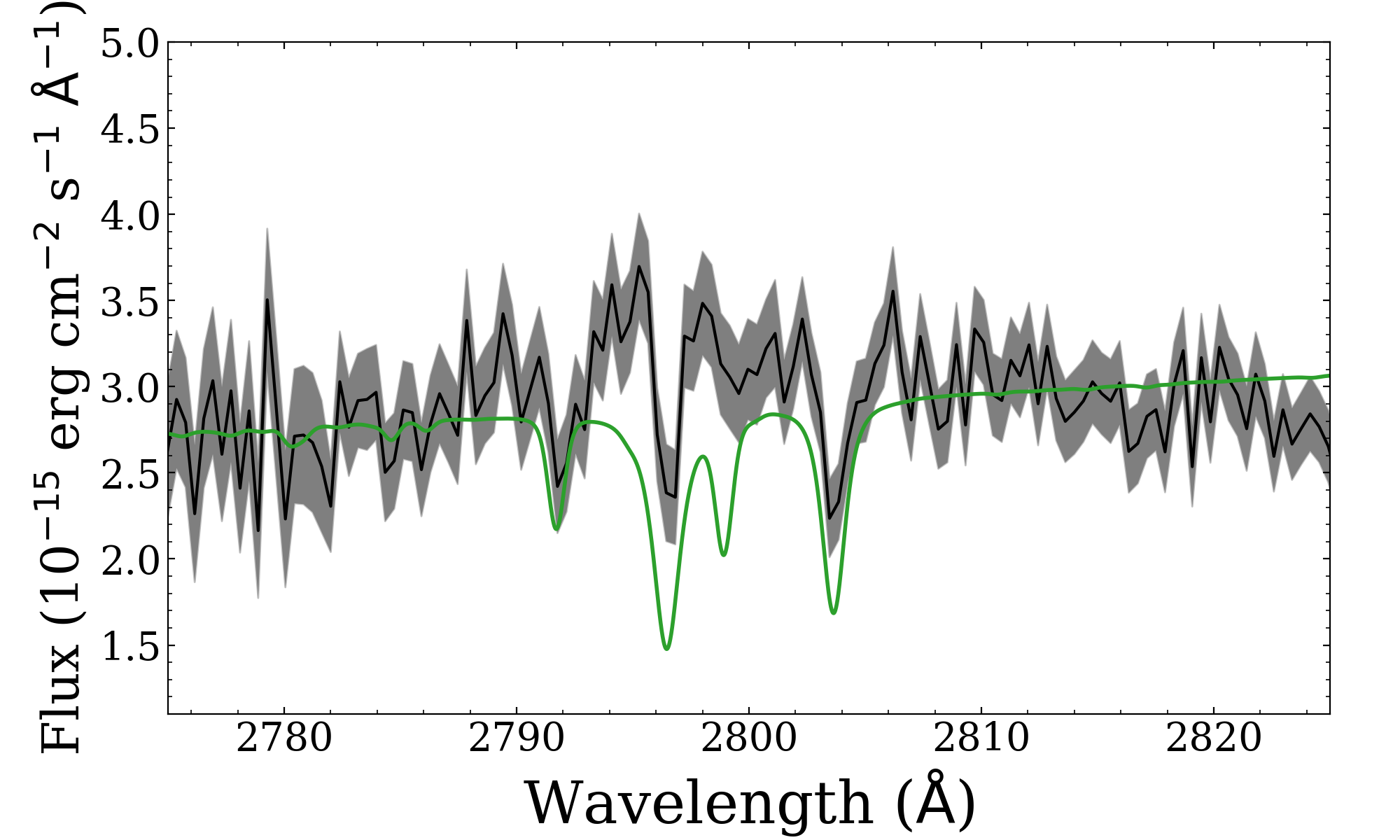}
    \caption{A zoomed-in view around the Mg II 2796~{\AA} region. The black line is the data with the grey area {\color{black} indicating} the uncertainty. The green line is the best fit atmospheric model, where photospheric absorptions from Mg II are detected \citep{Rogers2024}. There are additional broad Mg II emission {\color{black} lines} from the circumstellar gas.
    }
\label{fig:Mg2800}
\end{figure}

We performed an independent emission line analysis of Gaia~J0611$-$6931 using all available spectra. The identifications for the optical spectra are straightforward because most emission lines are resolved and not blended with other features. There is one triple-peaked feature that extends between 5160~{\AA} to 5190~{\AA} and it has been previously attributed to a blend of Fe II 5170~{\AA} and the Mg I triplet at 5169, 5174 and 5185~{\AA} \citep{Melis2020,Dennihy2020b}. As discussed in Section~\ref{sec: model II}, we think this feature is dominated by the Mg I triplet with little contribution from Fe II. The centroid of the Mg I 8809 {\AA} line is shifted by about $-90~km~s^{-1}$ compared to all the other emission lines. It is unclear what the cause of this is. There may be a similar wavelength shift for the Mg I 3832 {\AA} line too, though the emission line is a lot weaker, making it difficult to confirm. In the NUV data, there is a broad emission feature around 2800~{\AA} that comes from four Mg I lines {\color{black} (see Figure~\ref{fig:Mg2800})}. The infrared line identification {\color{black} in the Flamingos-2 data} is more challenging because many lines are blended, and our result is consistent with those reported in \citet{Owens2023}. 

To characterize the emission features, we measured the Full Width at Zero Intensity (FWZI) and the Equivalent Width (EW). For the EW, the uncertainty is dominated by the choice of continuum. For each line, we did several measurements selecting different wavelength regions for estimating the continuum flux. We report the average and the standard deviation of the measurements as the EW and its uncertainty, respectively. We also assume a 3\% minimum uncertainty in the EW of the optical lines given the S/N of the spectra. The Si I 3907~{\AA}, {\color{black} Si I 4104~{\AA},} and Ca I 3935~{\AA} lines sit on the slope of the broad hydrogen lines, and therefore have large uncertainties in the EW. The Ca II 3970~{\AA} line is also detected but it is difficult to perform any measurements due to its proximity to H$\epsilon$ 3971~{\AA}. The EWs and FWZIs of the infrared lines are taken from \citet{Owens2023}, which used the Flamingos-2 data. All the measurements are reported in Table~\ref{tab:lines}. 

\begin{deluxetable*}{lcccccccccc}
\tablecaption{ \label{tab:lines} Emission Line Measurements.}
\tablehead{
\colhead{Transition} &  \colhead{FWZI } & \colhead{EW}& \colhead{Line Flux} & \colhead{Other} & \colhead{Reference} \\
(vacuum {\AA})& \colhead{(km s$^{-1}$)} & \colhead{ ({\AA})} & \colhead{(10$^\mathrm{-16}$ erg cm$^\mathrm{-2}$ s$^\mathrm{-1}$)}
}
\startdata
O I 7774.1\tablenotemark{$\dag$} &1647$\pm$351 & 1.95$\pm$0.06 & 4.8$\pm$0.2 & O I 7776.3, 7777.5 & {\it This Paper}\\
O I 8448.6\tablenotemark{$\dag$} &1664$\pm$387 & 2.25$\pm$0.07& 4.1$\pm$0.2 & O I 8448.7 & {\it This Paper}\\
Na I 5891.6\tablenotemark{$\dag$} & 1740$\pm$479& 1.96$\pm$0.09 & 12.3$\pm$0.6 & Na I 5897.6& {\it This Paper} \\
Mg I 3833.4 &1361$\pm$63 & 0.26$\pm$0.02& 4.1$\pm$0.2 &... & {\it This Paper}\\
Mg I 5168.8\tablenotemark{$\dag$}& 2682$\pm$633 & 4.50$\pm$0.13 &45.2$\pm$1.4  &Mg I 5174.1, 5185.1 & {\it This Paper}  \\ 
Mg I 8809.2 &1399$\pm$201 & 6.77$\pm$0.20 & 11.1$\pm$0.4 & ... & {\it This Paper} \\
Mg I 11831\tablenotemark{$*$} &3226$\pm$560 & 10$\pm$2 & 5.4$\pm$0.7& ... & \citet{Owens2023} \\
Mg I 14882\tablenotemark{$\dag$} & 3558$\pm$447&15$\pm$2 &2.1$\pm$0.2 &five lines around Mg I 14882 & \citet{Owens2023} \\ 
Mg I 15029\tablenotemark{$\dag$} &2400$\pm$122  & 40$\pm$3 & 13.5$\pm$1.4& Mg I 15044, 15052 &\citet{Owens2023} \\
Mg I 17113 & 1928$\pm$363& 5.8$\pm$1.4 &1.9$\pm$0.2 &... & \citet{Owens2023} \\
Mg II 2796.4\tablenotemark{$\dag$} & 3530$\pm$600 & 1.2$\pm$0.3 & 92$\pm$10\tablenotemark{$\ddag$} & Mg II 2791.6, 2798.8, 2803.5 &{\it This Paper} \\
Si I 3906.6 & 1597$\pm$369& 1.80$\pm$0.14 & 27.2$\pm$2.0 &  ... & {\it This Paper}\\
{\color{black} Si I 4104.1 }& {\color{black} ...} &{\color{black} ...} & {\color{black} ...} &{\color{black} ...} & {\color{black} {\it This Paper}} \\
Si I 11987\tablenotemark{$\dag$} & 5800$\pm$600& 34.6$\pm$3.6 & 15.9$\pm$1.6 & Si I 11994,12034,12107,Mg I 12086  &\citet{Owens2023} \\
Si I 12274 &2165$\pm$266 & 4.9$\pm$0.8 &1.6$\pm$0.2 &... &\citet{Owens2023} \\
Si I \& Mg I \tablenotemark{$\dag$}  & 5900$\pm$600&36$\pm$3 &9.6$\pm$1.0 &Si I 15893, 15837, 15964 & \\ 
 & & & & Mg I 15770,15745,15753  &\citet{Owens2023}  \\
Si I 16385\tablenotemark{$\dag$} & 3697$\pm$396& 9.2$\pm$1.2 &1.8$\pm$0.2 &Si I 16386 & \citet{Owens2023} \\
Ca II 3934.8 & 1768$\pm$381 & 1.89$\pm$0.38 & 30.6$\pm$5.8 & ... &  {\it This Paper}\\
Ca II 3969.6  &... &... & ... &... &  {\it This Paper}\\
Ca II 8500.4 & 1439$\pm$148 & 13.15$\pm$0.40& 24.1$\pm$0.8 &... &  {\it This Paper}\\
Ca II 8544.4 & 1470$\pm$207  & 18.20$\pm$0.55&  32.8$\pm$1.0& ... &  {\it This Paper}\\
Ca II 8664.5 & 1530$\pm$284& 16.70$\pm$0.60& 28.9$\pm$0.9 & ... &  {\it This Paper}\\
  \enddata
\tablenotetext{$\dag$}{This line is a blend and only the strongest transition is listed. The adjacent lines that also contributed to the measurements are listed in the ``Other" column.}
\tablenotetext{$*$}{\color{black} Possibly blended due to the large FWZI.}
\tablenotetext{$\ddag$}{From the data, the measured line flux is 34$\pm$8$\times$10$^\mathrm{-16}$ erg cm$^\mathrm{-2}$. However, as shown in Figure~\ref{fig:Mg2800}, there is significant contribution from the photospheric absorption lines. The line flux of the photospheric lines are measured to be 58$\pm$6$\times$10$^\mathrm{-16}$ erg cm$^\mathrm{-2}$ from the model. The final number here is the summation of these two values. 
}
\end{deluxetable*}

We also measure the integrated line flux of all the emission lines, which is the flux above the continuum in absolute flux units. Note that in some cases, the photospheric absorption is visible in the profile, such as Ca II~3935~{\AA} and O I~8488~{\AA}. However, their overall contribution is small and are therefore not subtracted out in the measurement of line flux. In the case of the Mg II 2796~{\AA}, the contribution from the photospheric line is significant, as shown in Figure~\ref{fig:Mg2800}. The integrated line flux of the Mg II absorption lines {\color{black} from the model spectrum} was added back to derive the flux of the emission feature around Mg II 2976~{\AA}. Similar to the EW measurements, the uncertainty of line flux is dominated by the choice of continuum. For each line, we did several measurements selecting different wavelength ranges given the uncertainties in the FWZIs. We also assume a 3\% minimum uncertainty in the line flux for the optical lines and a 10\% minimum uncertainty for the infrared lines based on the S/N of the data.

As discussed in \citet{Gentile-Fusillo2021a}, the EW of an emission line is a relative number, which is normalized by the continuum flux. In comparison, the line flux is an absolute number above the continuum flux, which is a more intrinsic way to characterize the emission line strength. For example, Table~\ref{tab:lines} shows that the EW of the Ca II 3934~{\AA} line is approximately 10 times weaker than that of the Ca II 8544~{\AA} line. However, that is not due to the relative strength of the emission line, but the difference in the white dwarf continuum flux. As shown in Figure~\ref{fig:spectra_all}, the white dwarf continuum flux is about a factor of 10 higher around Ca II 3934~{\AA} than the Ca infrared triplet region. Therefore, the line strengths for Ca II 3934~{\AA} and Ca II 8544~{\AA} lines are actually similar. For the remainder of the manuscript, we will use line flux as the main observable to compare with the model outputs.

\begin{figure}[tbh]
    \centering
    \includegraphics[width=0.45\textwidth]{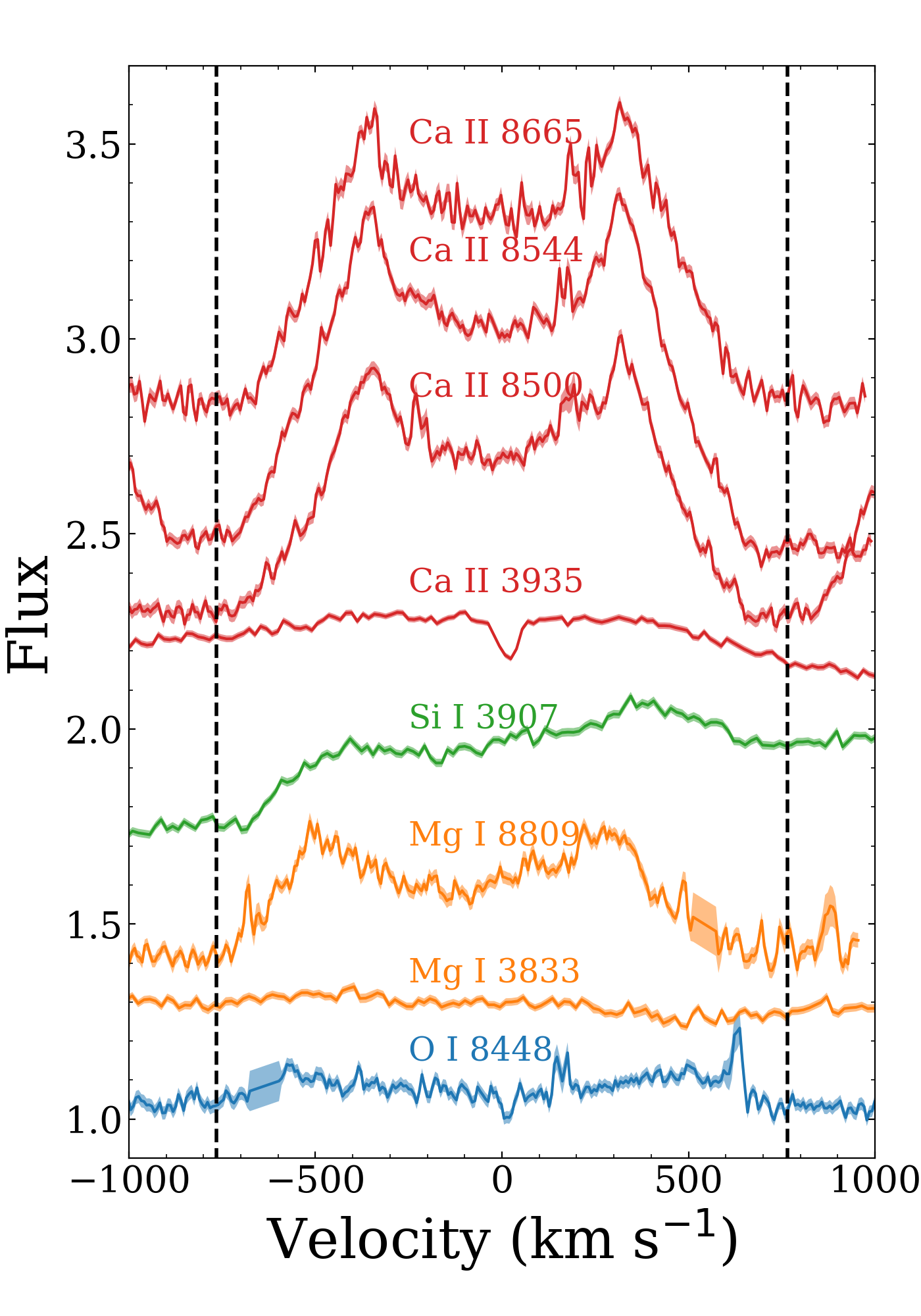}
    \caption{A compilation of all the unblended and resolved emission features listed in Table~\ref{tab:lines}. The O I 8448 doublet is also included because their separation is only 0.1~{\AA}. {\color{black} For each emission line, the continuum flux is normalized to one, and then offset for clarity. The velocity is shown in the reference frame of the white dwarf, and therefore the narrow photospheric absorption lines in Ca II 3935~{\AA} and O I 8448~{\AA} appears to be at 0 km s$^{-1}$.} The black dashed line marks a FWZI of 1530~km s$^{-1}$, which is the average {\color{black} width} of these lines.
    }
\label{fig:emission_all}
\end{figure}

Figure~\ref{fig:emission_all} shows a compilation of all the unblended and resolved emission features. The overall shape of the lines is a bit different from element to element, but they all have a similar FWZI of 1530~km s$^{-1}$, half of which corresponds to a Keplerian velocity at a radius of 20$\mathrm{R_\mathrm{WD}}$ assuming the gas is in a circular orbit and edge-on. {\color{black} If the disk is inclined relative to our line of sight, the projected velocity would correspond to a smaller radius}. The half peak separation is about 350~km s$^{-1}$, which corresponds to a Keplerian velocity at a radius of 70$\mathrm{R_\mathrm{WD}}$. The photospheric component can be seen as the narrow absorption feature around 0 km s$^{-1}$ in Ca II 3935~{\AA} and O I doublet at 8448~{\AA}. 
 
\section{Circumstellar Disk Modeling \label{sec: method}}

\subsection{Cloudy Modeling}

Here, we explore modeling the emission lines around Gaia~J0611$-$6931 using the photoionzation code \textsc{Cloudy} \citep{Ferland2017}. The \textsc{Cloudy} version 17.03 was used for this work. The setup is very similar to those described in \citet{Steele2021}. Our configuration {\color{black} uses the cylinder geometry and} assumes that the circumstellar gas has a circular orbit, a constant density, and is edge-on. {\color{black} \textsc{Cloudy} is a 1D code and therefore it only computes the radiative transfer in the radial direction.} The central white dwarf is the only ionizing source, whose parameters are listed in Table~\ref{tab:properties}. The main free parameters for the \textsc{Cloudy} calculation are the radial extent of the disk, the hydrogen number density, and the abundances of different elements, {\color{black} the latter two affecting both the total mass of the gas and the excitation of the different elements}.

For \textsc{Cloudy} outputs, we saved the predicted line intensities for a list of lines, which can be directly compared to the measured line fluxes in Table~\ref{tab:lines}. The line emissivity file, which reports the emergent emissivity as a function of cloud depth, was also saved to calculate the line profile (see Section~\ref{sec: line profile}). Other output files were also saved following \citet{Steele2021}. 

\subsection{Line Profile Modeling \label{sec: line profile}}

\begin{figure}[tbh]
    \centering
    \includegraphics[width=0.45\textwidth]{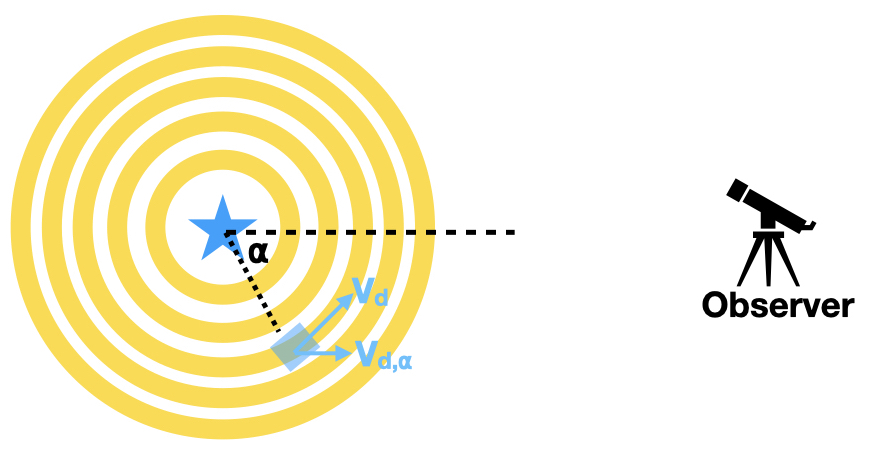}
    \caption{A top down view of the geometry of the gas disk. The white dwarf is the blue star and the gas disk is modelled as circular rings. For each small patch in the disk, the line of sight velocity V$_\mathrm{d,\alpha}$ can be directly computed and the emissivity can be obtained from \textsc{Cloudy}. \label{fig: illustration}
    }

\end{figure}

We adopted an analytical method to compute the line profile, as illustrated in Figure~\ref{fig: illustration}. An additional assumption here is that the gas disk is optically thin, therefore we can see the radiation all the way to the inner radius of the disk. As shown in Section~\ref{sec: model II}, this assumption is not supported by the \textsc{Cloudy} calculations. Using polar coordinates, the disk can be approximated as many small patches along the radial and the azimuthal directions. The Keplerian velocity $V_\mathrm{d}$ at a given distance d from the white dwarf is:

\begin{equation}
V_\mathrm{d}= \sqrt{\frac{GM_\mathrm{wd}}{d}} = 753\, \mathrm{km}\, \mathrm{s}^{-1} \sqrt{\frac{20R_\mathrm{wd}}{d} }
\end{equation}
where G is the gravitational constant; M$_\mathrm{wd}$ and R$_\mathrm{wd}$ is the mass and radius of the white dwarf. The velocity along the line of sight V$_\mathrm{d,\alpha}$ = V $\times$ sin $\alpha$, where $\alpha$ is the azimuthal angle of the patch.

For each patch, the emissivity is $\epsilon$(d), which is a function of the distance from the white dwarf. The emissivity is also a direct output from \textsc{Cloudy}. The line profile of the emission line is calculated as the sum of the emissivity of the patch in both the azimuthal and the radial directions $\Sigma_{d, \alpha}$ $\epsilon$(d), which is a function of the line of sight velocity (or wavelength).

\section{Result \label{sec: result}}

\subsection{Radial Extent of the Disk \label{sec: model I}}

In this section, our goal is to {\color{black} use \textsc{Cloudy} to better constrain the radial extent of the disk}, i.e., the inner disk radius R$_\mathrm{in}$ and the outer disk radius R$_\mathrm{out}$. The gas disk is estimated to extend from 20R$_\mathrm{wd}$ to 70R$_\mathrm{wd}$ based on the profiles of the double peaked emission lines (see Section~\ref{sec: line analysis}). 

The hydrogen number density is another basic input for \textsc{Cloudy}, but this value is difficult to constrain because hydrogen emission lines are not detected. {\color{black} For the ease of the calculation, we fix the ratio between hydrogen and calcium to be 1 (by number), because it is possible to estimate the density of calcium.} Assuming a constant density profile, we can calculate the number density of calcium n(Ca) as,

\begin{equation}
n(Ca)= \cfrac{M_\mathrm{Ca, gas}}{ u_\mathrm{Ca}} \times \cfrac{1}{\pi (100R_\mathrm{wd})^2 \times h } 
\label{equ:nCa}
\end{equation}
where M$_\mathrm{Ca,gas}$ is the mass of calcium {\color{black} in the circumstellar gas}, $u_\mathrm{Ca}$ is the atomic mass of calcium, and $\pi (100R_\mathrm{wd})^2 \times {\color{black} h}$ is the estimated volume for the gas, assuming a constant disk scale height {\color{black} h} of 1~$R_\mathrm{wd}$. {\color{black} M$_\mathrm{Ca,gas}$ can be estimated from the mass of Ca in the white dwarf's photosphere M$_\mathrm{Ca,phot}$ (7$\times$10$^\mathrm{11}$ g from \citealt{Rogers2024}), the settling time of Ca in the white dwarf's photosphere $\tau_\mathrm{Ca}$ (10$^{-2.539}$ yr from \citealt{Dufour2016}), and the accretion timescale of the disk $\tau_\mathrm{diff}$,}

\begin{equation}
{\color{black}
M_\mathrm{Ca,gas} = \cfrac{M_\mathrm{Ca,phot}}{\tau_\mathrm{Ca}} \times \tau_\mathrm{diff}
}    
\end{equation}

{\color{black} Using the $\alpha$ disk model, $\tau_\mathrm{diff}$ can be estimated using Equation (14) of \citet{Goksu2023}.} Putting in all the numbers, the calcium number density is,

\begin{equation}
{\color{black} 
n(Ca) = 4.2 \times 10^7\, \mathrm{cm}^\mathrm{-3} \times \cfrac{1R_\mathrm{wd}}{h} \times \cfrac{10^{-2}}{\alpha} 
}
\end{equation}

{\color{black} The viscosity parameter $\alpha$ is uncertain by a few orders of magnitude. Therefore, we explore the calcium number density n(Ca) over a range of values. 
}

{\color{black} We ran a few \textsc{Cloudy} models using different {\color{black} values} of disk scale height and found it is {\color{black} proportional} to the absolute strength of the emission lines. A larger scale height means a larger emitting area, and therefore a stronger emission line flux. For {\color{black} this work}, the disk scale height is kept at 1~$R_\mathrm{wd}$. Note that 1~$R_\mathrm{wd}$ is just a convenient value to adapt; it is {\color{black} larger} than the hydrostatic value, which depends on the disk temperature and radius \citep{Goksu2023}. }

\begin{figure}[tb]
    \centering
    \includegraphics[width=0.49\textwidth]{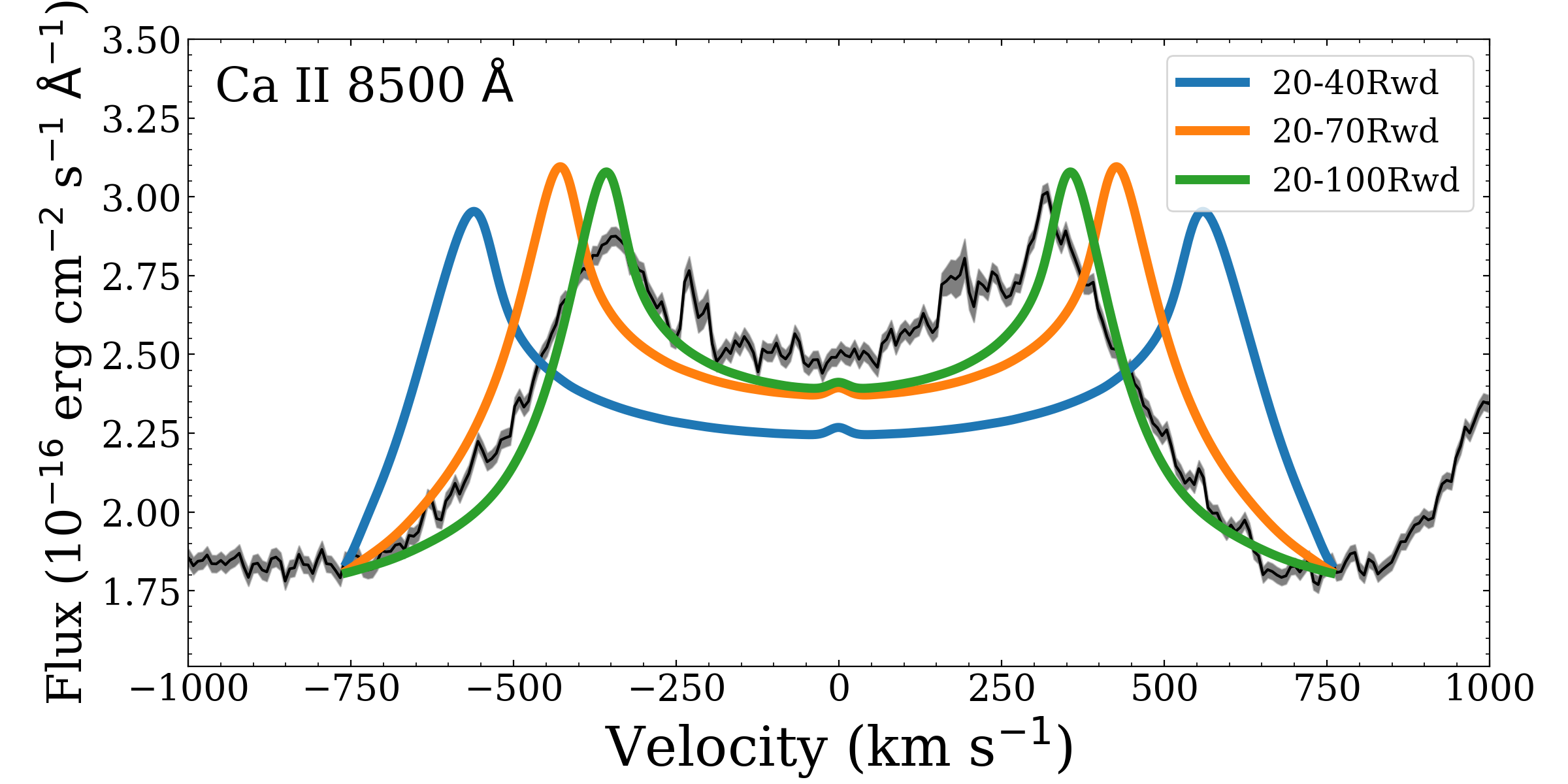}
    \includegraphics[width=0.49\textwidth]{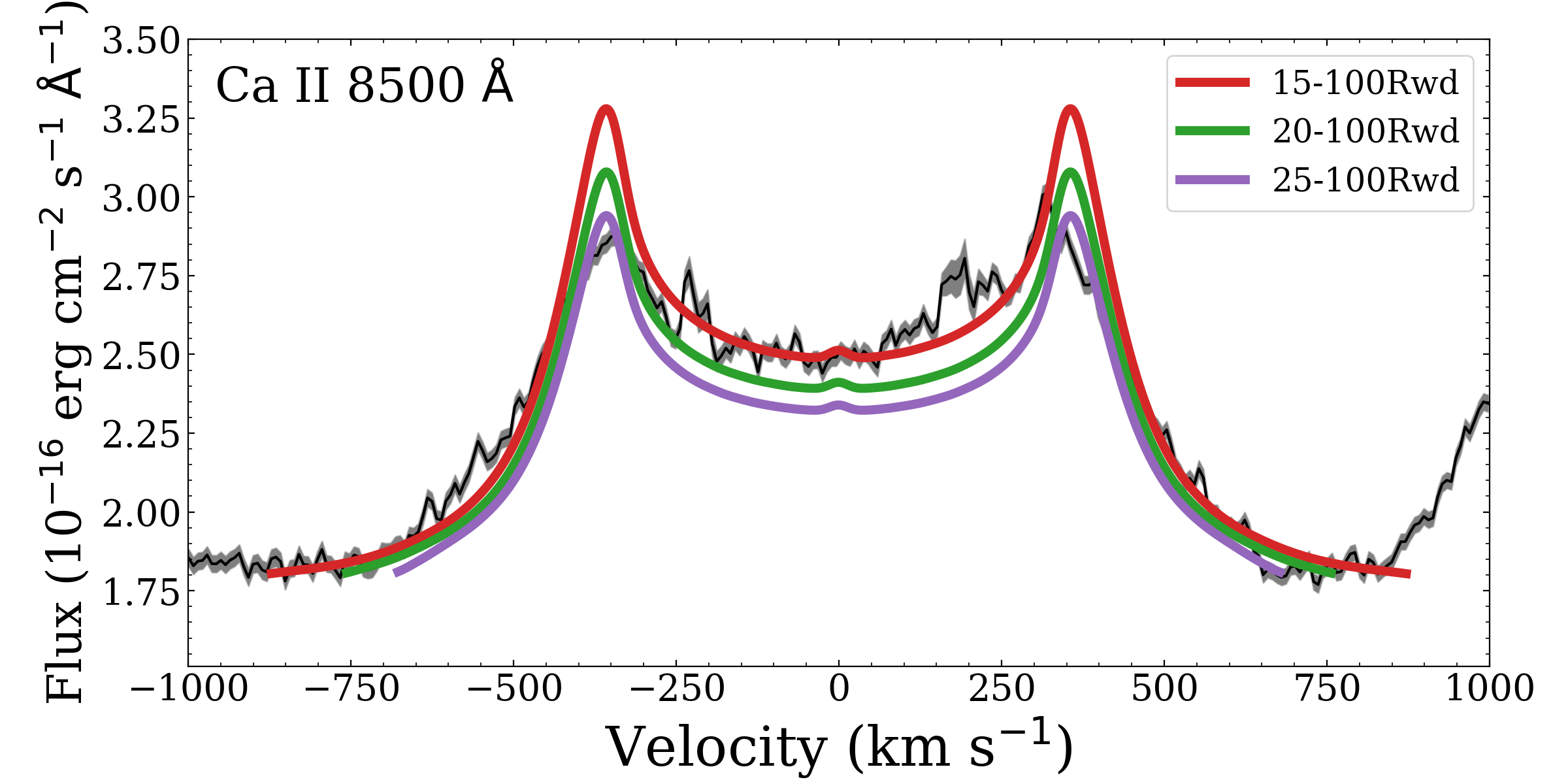}
    \caption{A comparison between the observed Ca II 8500~{\AA} line (black line with grey shade representing the uncertainties) and the models with different radial extents of the disk (colored lines). In the top panel, the models were scaled differently to match the overall emission line flux. In the bottom panel, the scaling factors are the same for all the models. {\color{black} Under the assumptions that the gas is optically thin with a constant density, the shape of the line profile is very sensitive to the outer radius of the gas. The model that best matches the observed line profile has a radial extent from 20R$_\mathrm{WD}$ to 100R$_\mathrm{WD}$.}}
\label{fig: profile}
\end{figure}

We compute a grid of \textsc{Cloudy} calculations with the inner radius R$_\mathrm{in}$ of 15R$_\mathrm{wd}$, 20R$_\mathrm{wd}$, 25R$_\mathrm{wd}$ and the outer radius R$_\mathrm{out}$ of 40R$_\mathrm{wd}$, 70R$_\mathrm{wd}$, and 100R$_\mathrm{wd}$. The result is shown in Figure~\ref{fig: profile}. The separation of the double peak is very sensitive to the outer radius of the disk, {\color{black} but not as much to the inner disk radius. This is a result of the assumed constant density profile of the gas, where most of the disk mass is in the outer part. Therefore, a different density profile will change the radial extent of the disk.} The model that best matches the line profile goes from 20R$_\mathrm{wd}$ to 100R$_\mathrm{wd}$.

\subsection{Disk Composition \label{sec: model II}}

Here, we {\color{black} attempt} to constrain the chemical composition of the circumstellar gas. We focus on the isolated lines listed in Table~\ref{tab:lines}. Mg I 11831~{\AA} is excluded from this analysis, because it is likely to be blended with other lines due to its large FWZI, even though a specific line has not been identified. For Oxygen, there are no isolated lines, so both O I triplet around 7774~{\AA} and the O I doublet at 8488~{\AA} are used. The disk is fixed at 20--100 R$_\mathrm{wd}$, the optimal parameters from Section~\ref{sec: model I}.

\begin{figure*}
\gridline{\fig{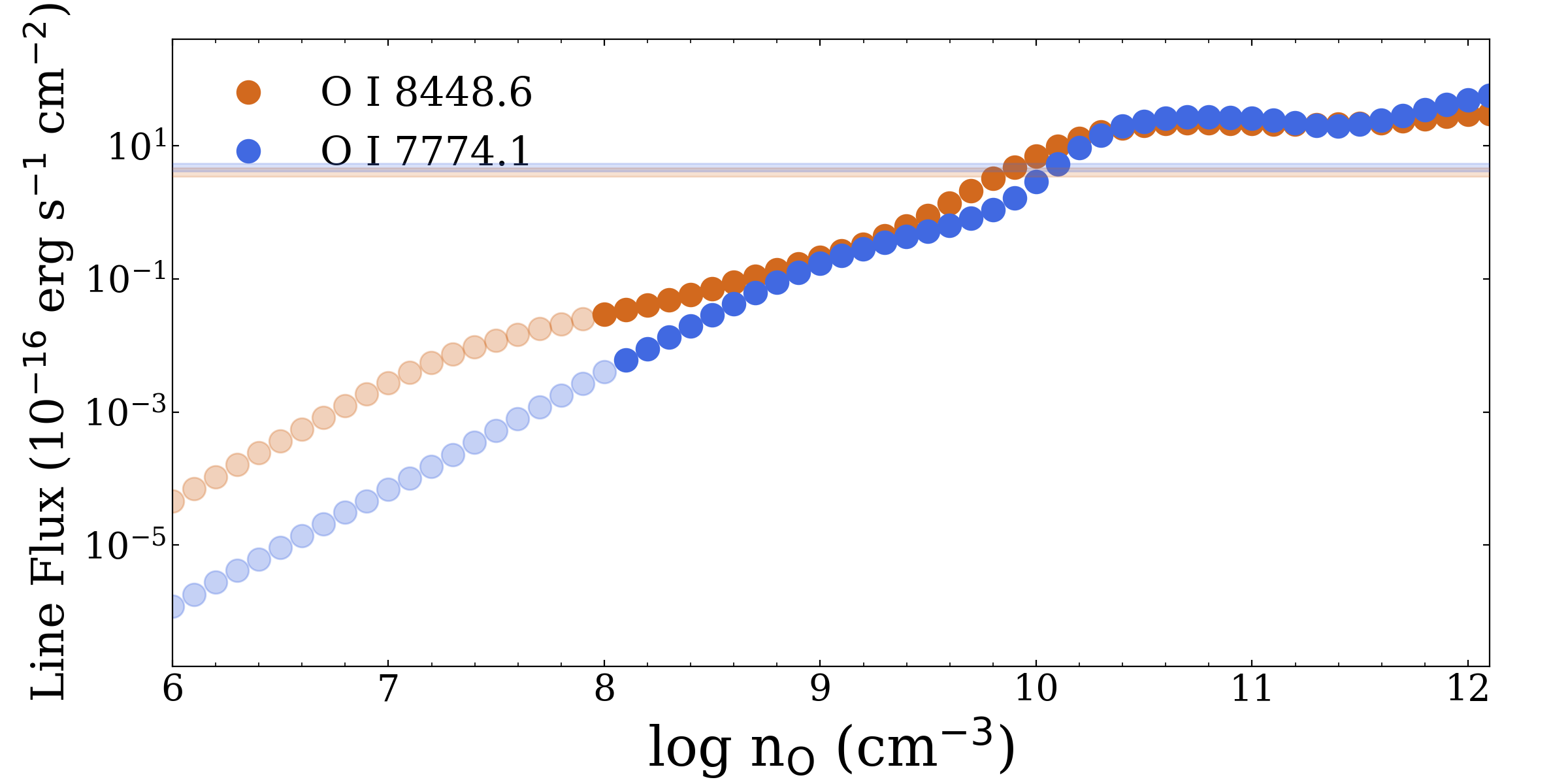}{0.5\textwidth}{}
\fig{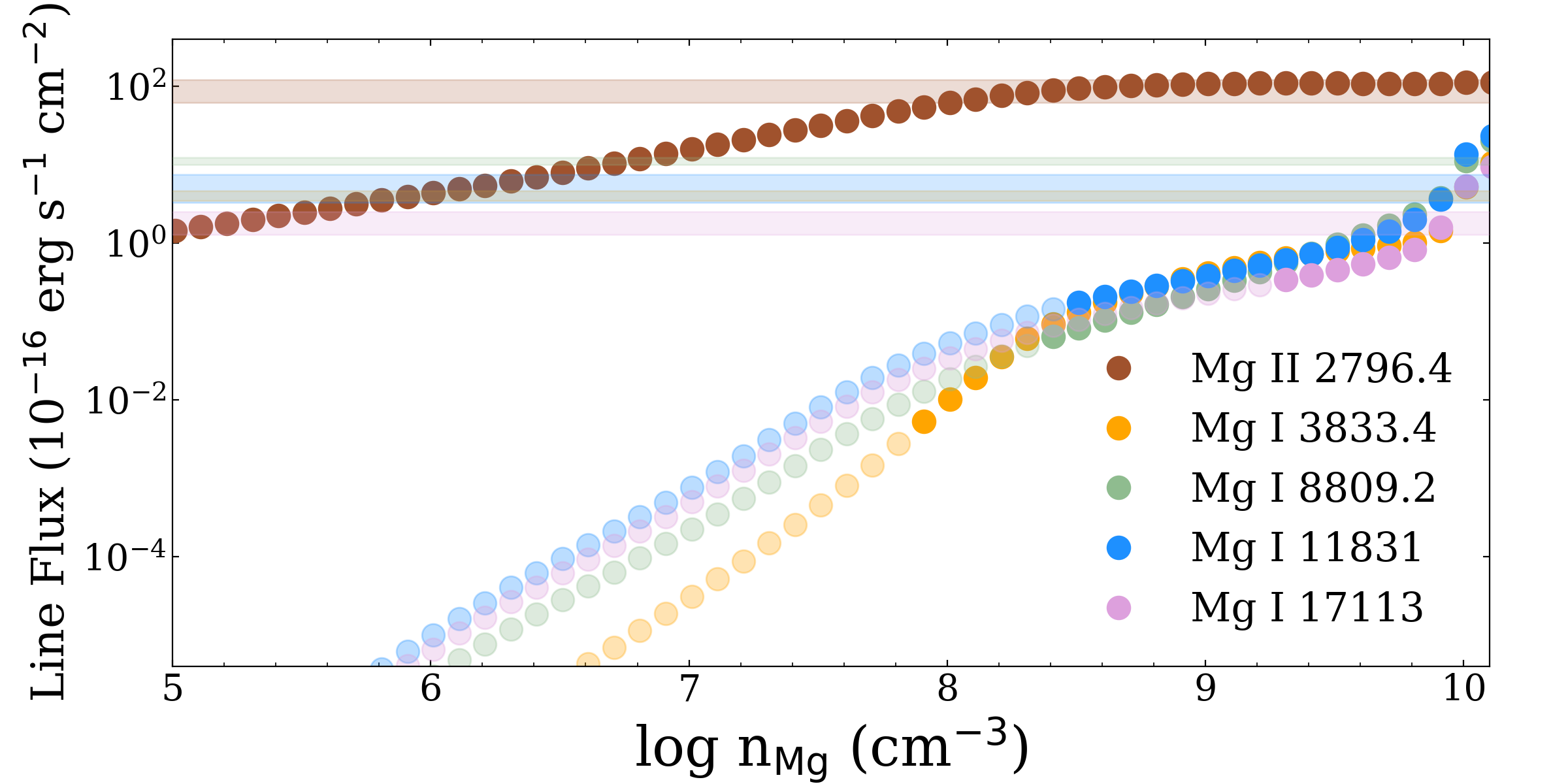}{0.5\textwidth}{}}
\gridline{\fig{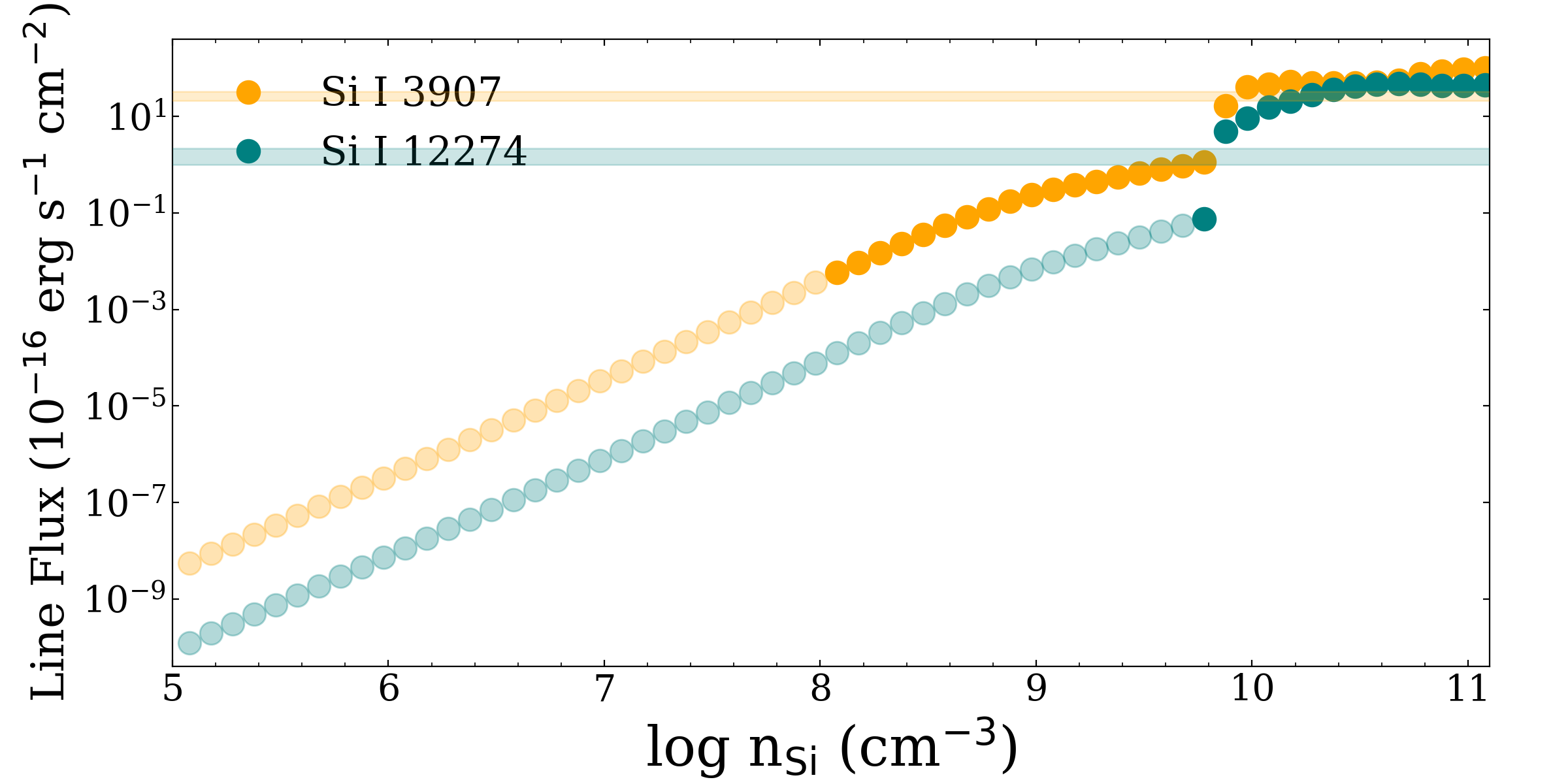}{0.5\textwidth}{}
\fig{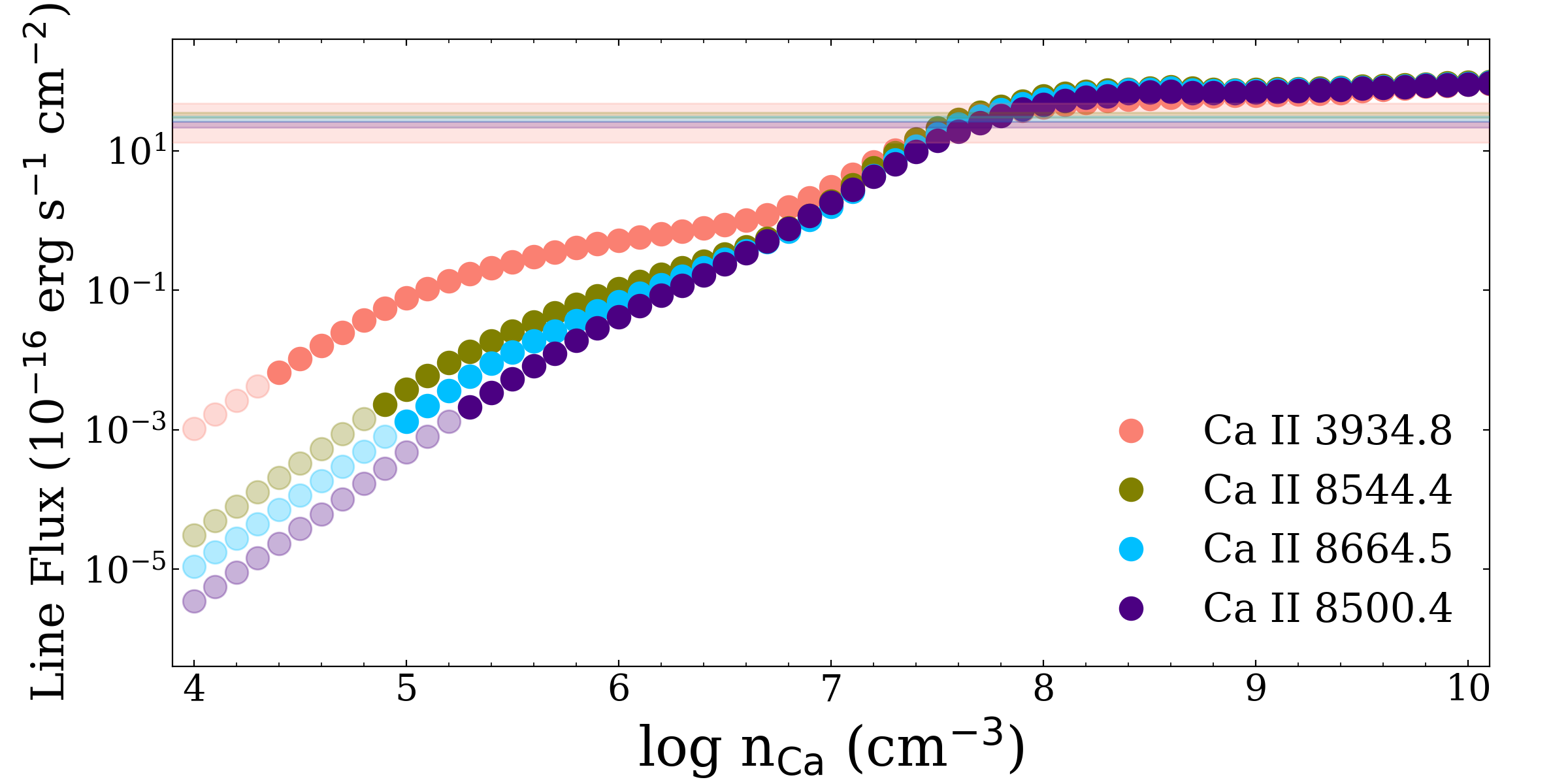}{0.5\textwidth}{}}
\centering
\includegraphics[width=0.49\textwidth]{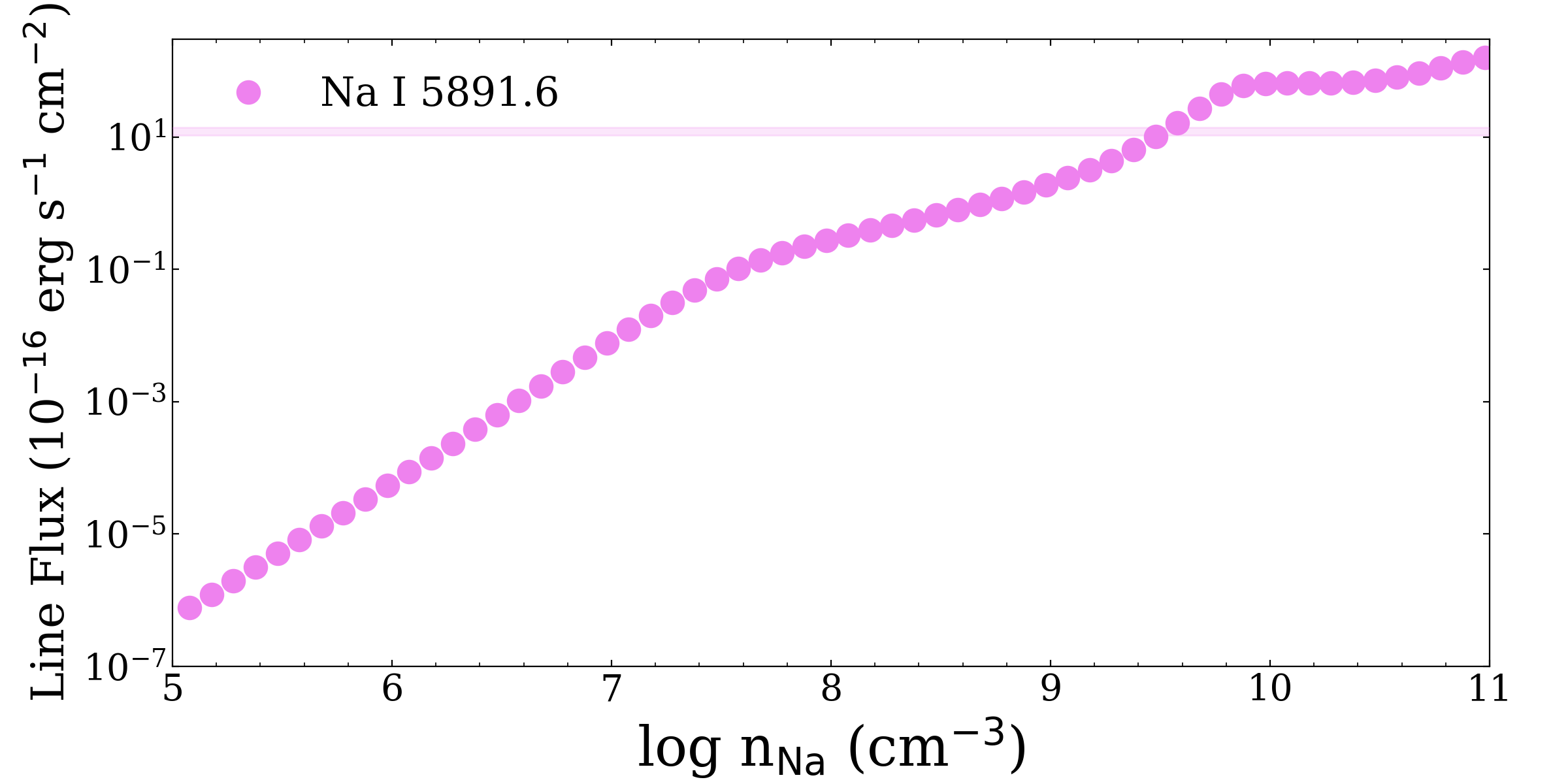}
\caption{  \textsc{Cloudy} models' prediction of {\color{black} the observed} line flux as a function of number density of each element ({\color{black} i.e., the curve of growth}). The optically thin regions are marked as transparent circles, {\color{black} while the optically thick regions are marked as solid circles}. The optical depth is measured radially. The shaded area represents the 3$\sigma$ range of the observed {\color{black} line flux} listed in Table~\ref{tab:lines}. {\color{black} Most of the emission lines appear to be optically thick.}
\label{fig: line flux ratio}}
\end{figure*}

For each element, we {\color{black} used \textsc{Cloudy} to compute the curve of growth, which shows the line flux as a function of the number density. In this iteration, we assumed the circumstellar gas has the same abundance as the bulk Earth, and varied the total amount of material}. The results are shown in Figure~\ref{fig: line flux ratio}. {\color{black} The lines that are mostly likely to be optically thin are the O I 7774 triplet, Mg I 17113~{\AA}, Si I 12274~{\AA}, and Ca II 8500.4~{\AA}}. However, given the observed line flux, most of the observed lines are optically thick.

\begin{figure*}
\gridline{\fig{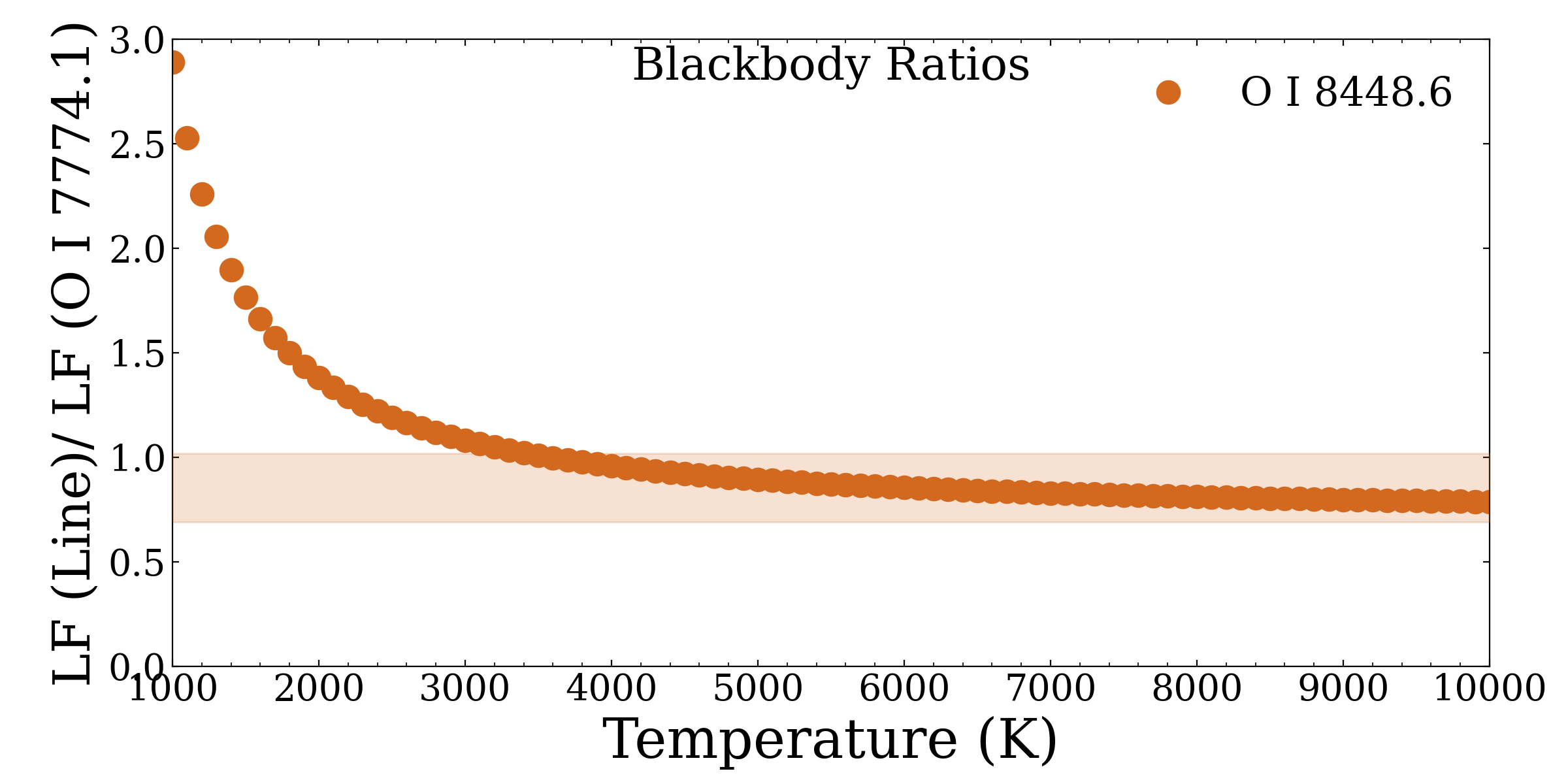}{0.5\textwidth}{}
\fig{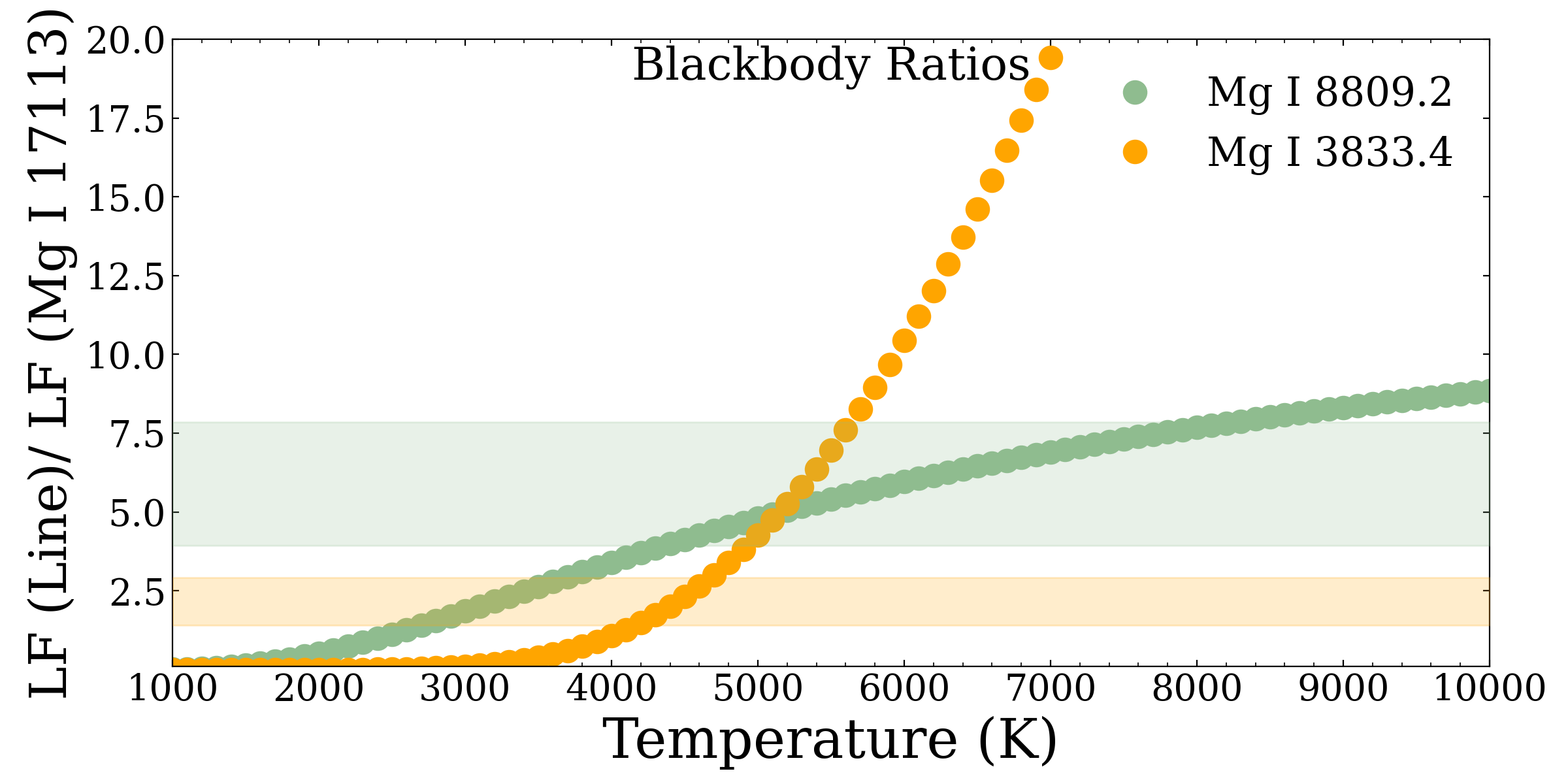}{0.5\textwidth}{}}
\gridline{\fig{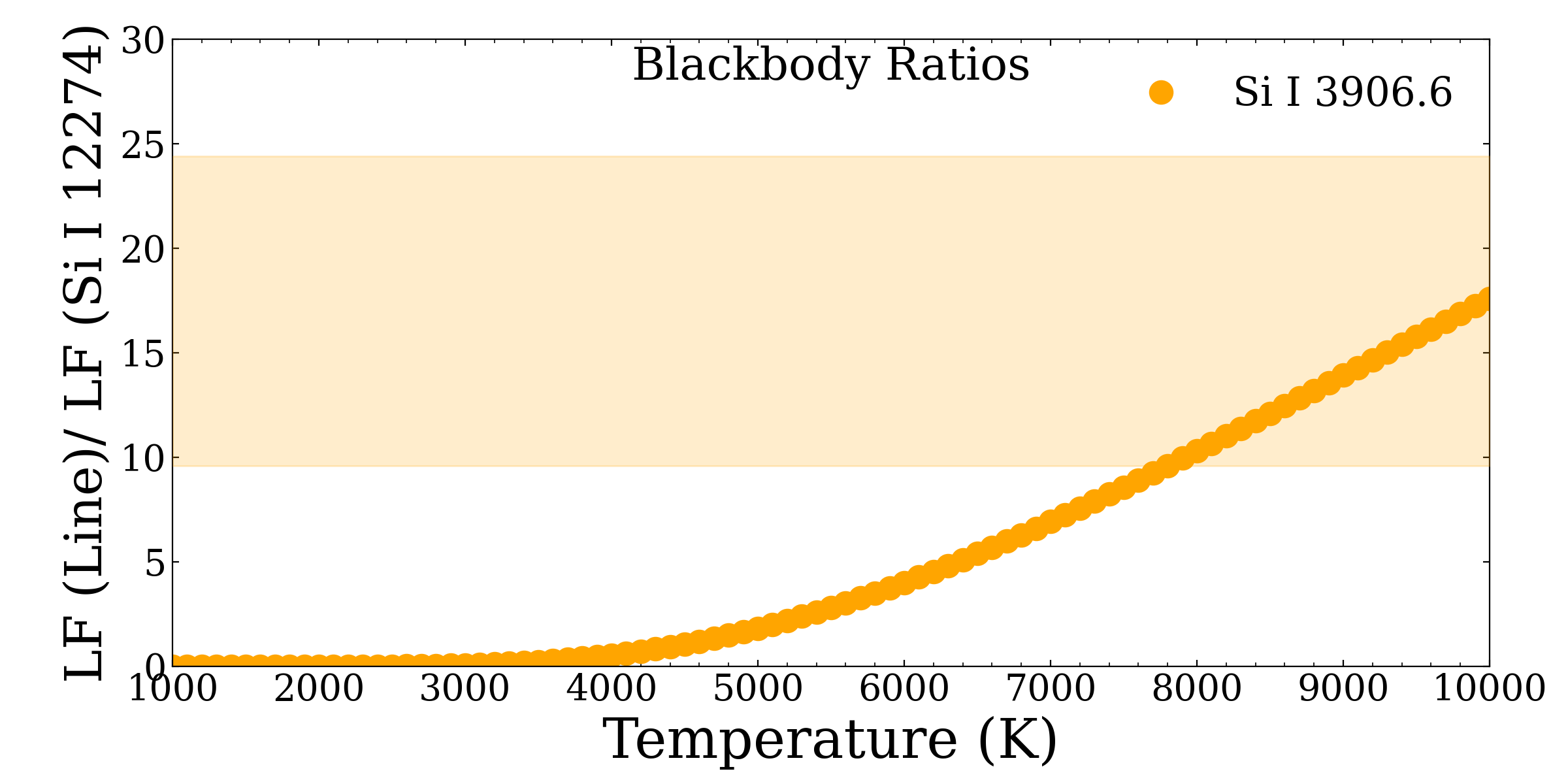}{0.5\textwidth}{}
\fig{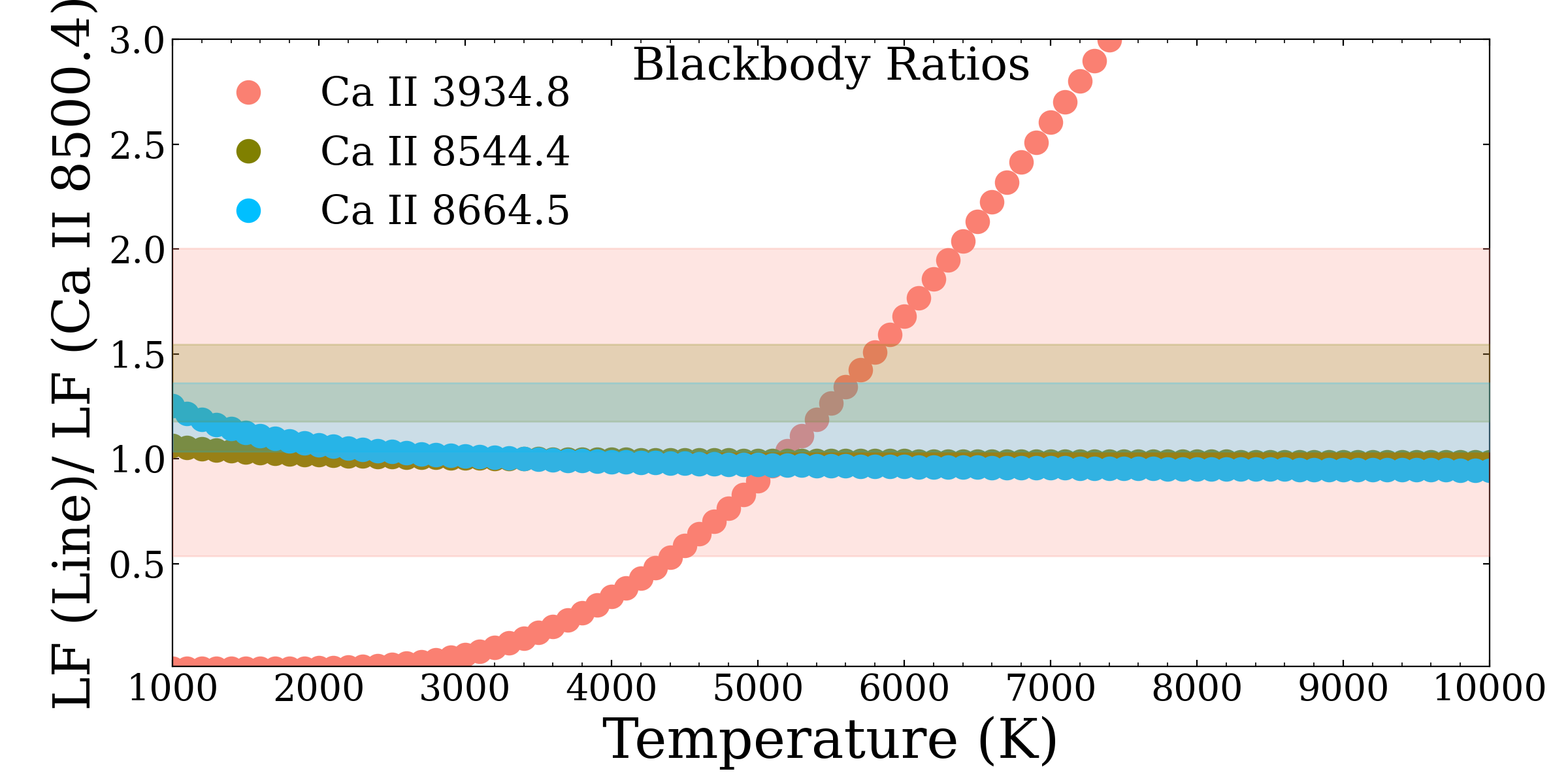}{0.5\textwidth}{}}
\caption{ {\color{black} The blackbody line ratios as a function of gas temperature. The shaded area represents 3$\sigma$ range of the observed line flux. With the possible exception of Si, all the other line ratios can be explained as the blackbody ratios. }
\label{fig: BB ratio}}
\end{figure*}

{\color{black} As an additional check, we consider the limiting case where the emission lines are completely saturated and the line flux reaches the Planck blackbody curve with a temperature equal to the gas temperature. In that case, the emission line ratios would equal to blackbody flux ratios. The result for this limiting case is presented in Figure~\ref{fig: BB ratio}. Within the uncertainties, the observed line ratios for O, Mg, and Ca all agree with the blackbody ratios for a temperature between 4,500--10,000~K (the computed gas temperature from \textsc{Cloudy}, see Section~\ref{sec: temperature}). For these elements, only an upper limit to the number density can be derived.} The only possible exception is Si. {\color{black} The Si I 12274~{\AA} can remain optically thin at a number density of 10$^{9.8}$ cm$^{-3}$, the highest of all the detected emission lines}. A silicon density may be derived using this line.

\begin{deluxetable}{lcccccccccc}[h!]
\tablecaption{ \label{tab:abundances} Circumstellar gas abundances from this work and the photospheric abundances from \citet{Rogers2024}, with spectroscopic parameters of the white dwarf and abundances from the optical spectra.}
\tablehead{
\colhead{Element} &  \colhead{Circumstellar } & \colhead{Photosphere} \\
& (log n(X) cm$^{-3}$) & (log n(X)/n(H)) \\
}
\startdata
O & {\color{black} $>$  9.7 } & -3.75 $\pm$ 0.16\\
Na & {\color{black} $>$ 6.8}  & - \\
Mg & {\color{black} $>$  9.0 }  & -4.61 $\pm$ 0.11 \\
Si & {\color{black} $>$ 10.3 }  & -4.70 $\pm$ 0.11\\
Ca & {\color{black} $>$  5.7 } & -6.08 $\pm$ 0.15 \\
  \enddata
\end{deluxetable}

One more iteration of the \textsc{Cloudy} model was computed {\color{black} to find the abundance that best match the observed line strengths. However, because the number densities of elements are similar, the abundances of one element can affect the line strength of the other element. More tweaks of the abundances are needed to find the best model. Our best fit abundance is listed in Table~\ref{tab:abundances}, which is able to reproduce the observed line strength within a factor of 2. All the lines have very large optical depths, and therefore only a lower limit can be derived for the density.
}The hydrogen number density is kept at {\color{black} 10$^{5.7}$} cm$^{-3}$, the same as the number density of calcium. The radial extent of the disk is fixed at 20--100R$_\mathrm{wd}$. As an additional confirmation, different \textsc{Cloudy} output files are checked to ensure that no other strong lines (such as hydrogen emission or forbidden lines) appear in the model. 

\begin{figure}
\includegraphics[width=0.49\textwidth]{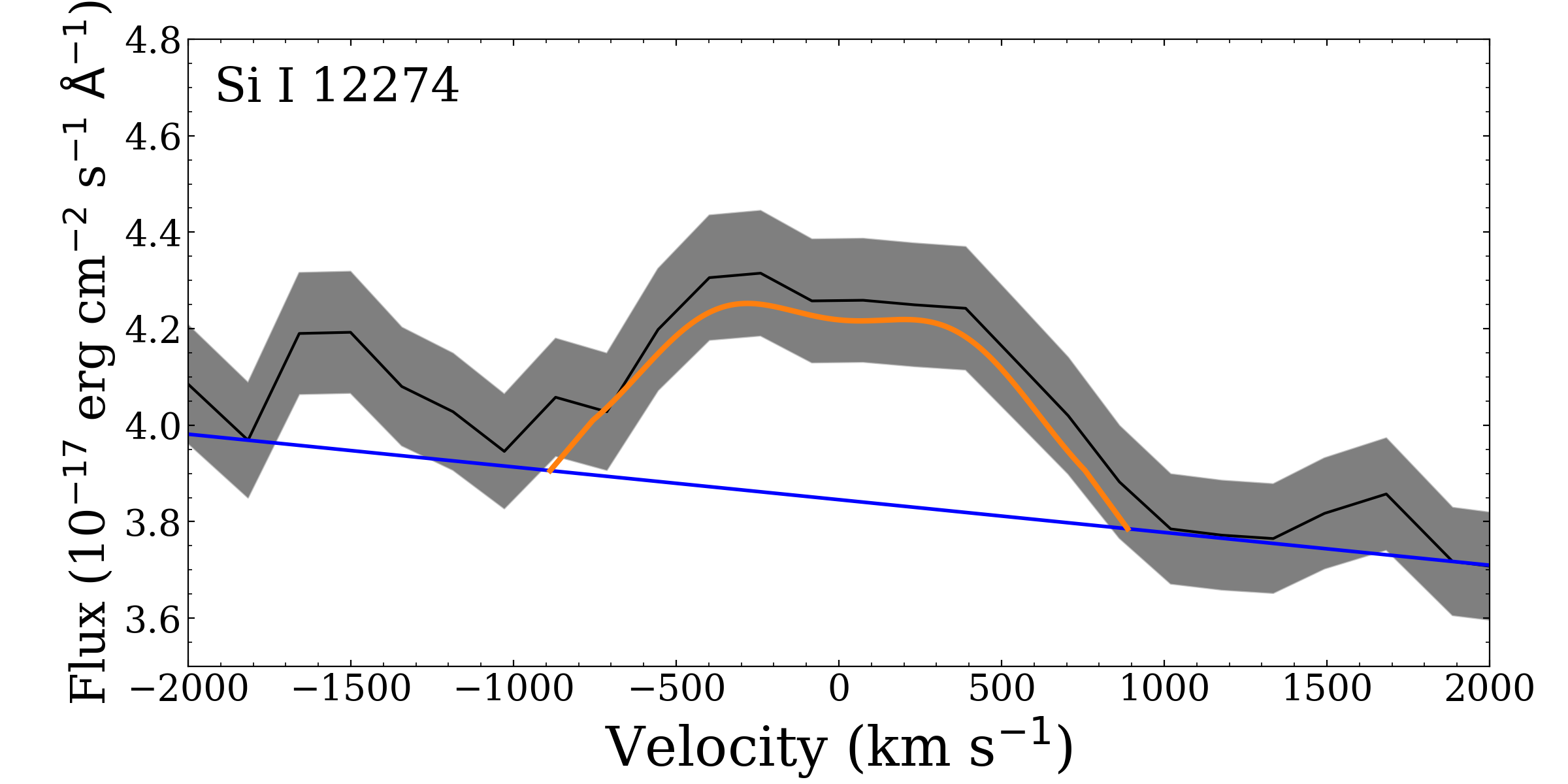}
\caption{A comparison between the observed line profile (black line with grey shaded region indicating 1$\sigma$ uncertainties) and our model prediction (orange line) for Si I 12274~{\AA}. \label{fig:Line Profile}}
\end{figure}

{\color{black} As a test to the line fitting code presented in Section~\ref{sec: line profile}, we computed a line profile for Si I 12274~{\AA} and compared it with data,} as shown in Figures~\ref{fig:Line Profile}. {\color{black} The match is respectable, but the spectral resolution of the data is too low to make a proper comparison. }

{\color{black} There is a three-peaked feature around 5170~{\AA}, which is a blend of three Mg I lines from 5168.8~{\AA}, 5174.1~{\AA}, 5185.1~{\AA}, and possibly Fe II 5170~{\AA} \citep{Dennihy2020b}. Our current model with only Mg already overpredicts the observed feature, and we conclude that the contribution from Fe II 5170~{\AA} in this feature should be minimal.
}

\subsection{Disk Temperature \label{sec: temperature}}

\begin{figure}[tbh]
    \centering
    \includegraphics[width=0.49\textwidth]{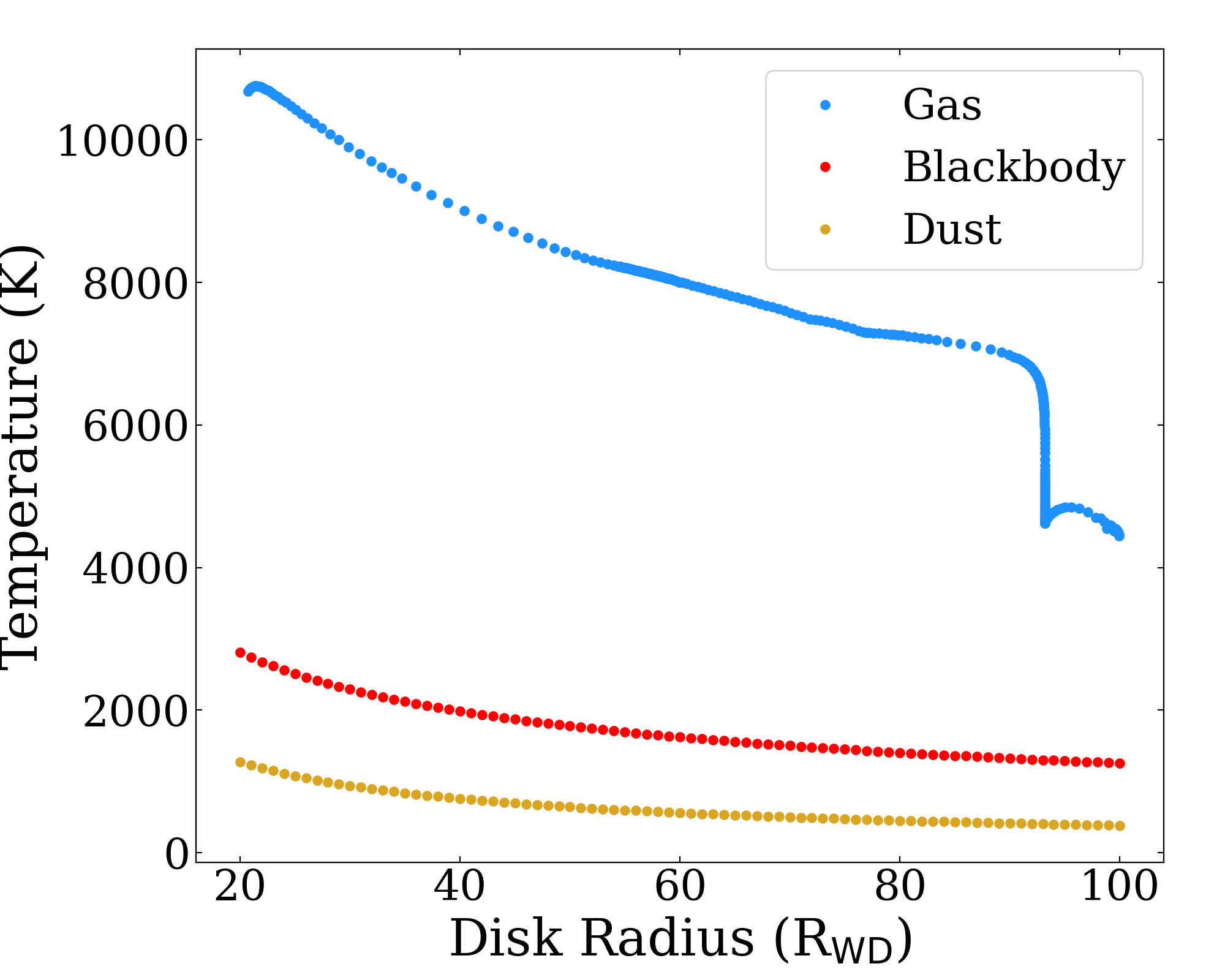}
    \caption{Gas temperature as a function of radius from \textsc{Cloudy} calculations (blue dots). The dust temperature using the model from \citet{Jura2003} (brown dots) and assuming a blackbody (red dots) is also shown for comparison. {\color{black} It is difficult for the dust and gas to co-exist, given the large temperature difference.}  }
\label{fig: temperature}
\end{figure}

The gas temperature profile from the best fit \textsc{Cloudy} model is shown in Figure~\ref{fig: temperature}. The temperature ranges from {\color{black} 10,000~K} around the innermost radius at 20~$R_\mathrm{WD}$ to {\color{black} 4,500~K} to the outer most radius at 100~$R_\mathrm{WD}$. The radiation from the white dwarf is the main heat source and cooling is from various emission lines. The gas temperature is similar to the analytical solution proposed by \citet{Melis2010}. They described a ``Z II" region model, which is akin to an H II region, and the gas is heated by ultraviolet light from the white dwarf and cools by optically thick emission lines. As mentioned in \citet{Steele2021}, \textsc{Cloudy} computes a self-consistent temperature profile, which is an advantage compared to other disk modeling papers assuming an isothermal disk \citep[e.g.][]{Fortin-Archambault2020,Budaj2022}.

Gaia~J0611$-$6931 has a bright infrared excess with a fractional luminosity of 5\% \citep{Dennihy2020b}. In fact, a flat disk model failed to reproduce such a strong infrared excess and the disk is either warped or has a significant amount of optically thin dust \citep{Owens2023}. Assuming there is dust around the same region as the circumstellar gas, we compute two dust temperatures using the optically thick model from \citet{Jura2003} and an optically thin model (assuming a blackbody). In both cases, the dust temperature is much cooler than the gas. {\color{black} It is difficult for the dust and gas to occupy the same region given the large temperature difference}. Alternatively, the dust and gas may not occupy the same region, as proposed for {\color{black} the eccentric disk around} SDSS~J1228+1040 \citep{Goksu2023}. Future work that incorporates both the dust and gas is needed to properly model these systems.

\subsection{Other Emission Lines}

\citet{Rogers2024} reported detections of photospheric C, S, P, Al, Fe and Ni in Gaia~J0611$-$6931, which were not detected in the circumstellar gas. We use the \textsc{Cloudy} model to make some predictions of the strongest emission lines from these elements, as listed in Table~\ref{tab:predicted Lines}. The lines are separated into UV, optical and infrared wavelengths. As discussed in Section~\ref{sec: line analysis}, generally speaking, the UV and optical line fluxes are larger compared to the infrared lines. However, the detectability of the emission lines are better in the infrared, because the white dwarf continuum flux is smaller. Effectively, it means it is possible to detect less luminous lines in the infrared. It will be useful to search for these additional emission lines to further constrain the composition of the circumstellar gas and compare with the photospheric abundances. A more extensive study exploring the detectability of the emission lines as a function of white dwarf and disk parameters will be presented in a future work (Steele in. prep).

\begin{deluxetable*}{llll}
\tablecaption{ \label{tab:predicted Lines} \textsc{Cloudy} Predicted Strong Emission Lines from Other Elements for Gaia~J0611$-$6931 }
\tablehead{
\colhead{Ions} & \multicolumn{3}{c}{Transition (vacuum {\AA}) } \\
& \colhead{UV (1000-3000 {\AA})} & \colhead{Optical (3000-10,000 {\AA})} &\colhead{Infrared (10,000-100,000 {\AA})}
}
\startdata
C I & 1561.5, 1656.2, 1658.1  &  4622.8, 8729.4, 9852.9 & 10710, 10757, 14547\\
C II & 1334.5, 1335.6, 1335.7 & 4746.1, 6584.8, 7233.3 & ... \\
S I & 1807.3, 1820.3, 1826.2 & 7727.1, 9230.7, 9240.2 & 10458, 10824, 11309\\
S II & 1102.4, 1250.6, 1253.8 & 4069.7, 4077.6, 6732.6 & 10290, 10323, 10339\\
P I & 1775.0, 1787.6,  2136.9 & 5334.0, 5341.1, 8789.9 & 13538, 13566, 13584 \\
P II & 1152.8, 1157.0, 1159.0 & 4670.5, 4737.9, 7878.2 & 11471, 11886, 11258\\
Al I & 1775.7, 2071.5, 2398.0 &  3083.0, 3093.6, 3962.6 & 13127, 13155, 16755 \\
Al II & 1670.8, 1725.0, 1763.9 & 3901.8, 7065.6 & 10126 \\
Fe I & 2181.0, 2485.0, 2527.3 & 3441.6, 3721.0, 3738.2 & 11611, 11887, 11976 \\
Fe II & 2396.3, 2612.7, 2626.5 & 3256.9, 3278.3, 5019.8 &12570, 13558, 14335 \\
Ni I & 1623.7 & 3036.6, 3311.9, 3481.1 & 31199, 58933, 75066\\
Ni II & 1324.1,1375.7, 1744.2 & 3439.9, 3627.9, 7379.9 & 10462, 19393, 66360\\
\enddata
\end{deluxetable*}

\section{Discussion \label{sec: discussion}}

\subsection{Composition Comparison}

\begin{figure}[tbh]
    \centering
    \includegraphics[width=0.49\textwidth]{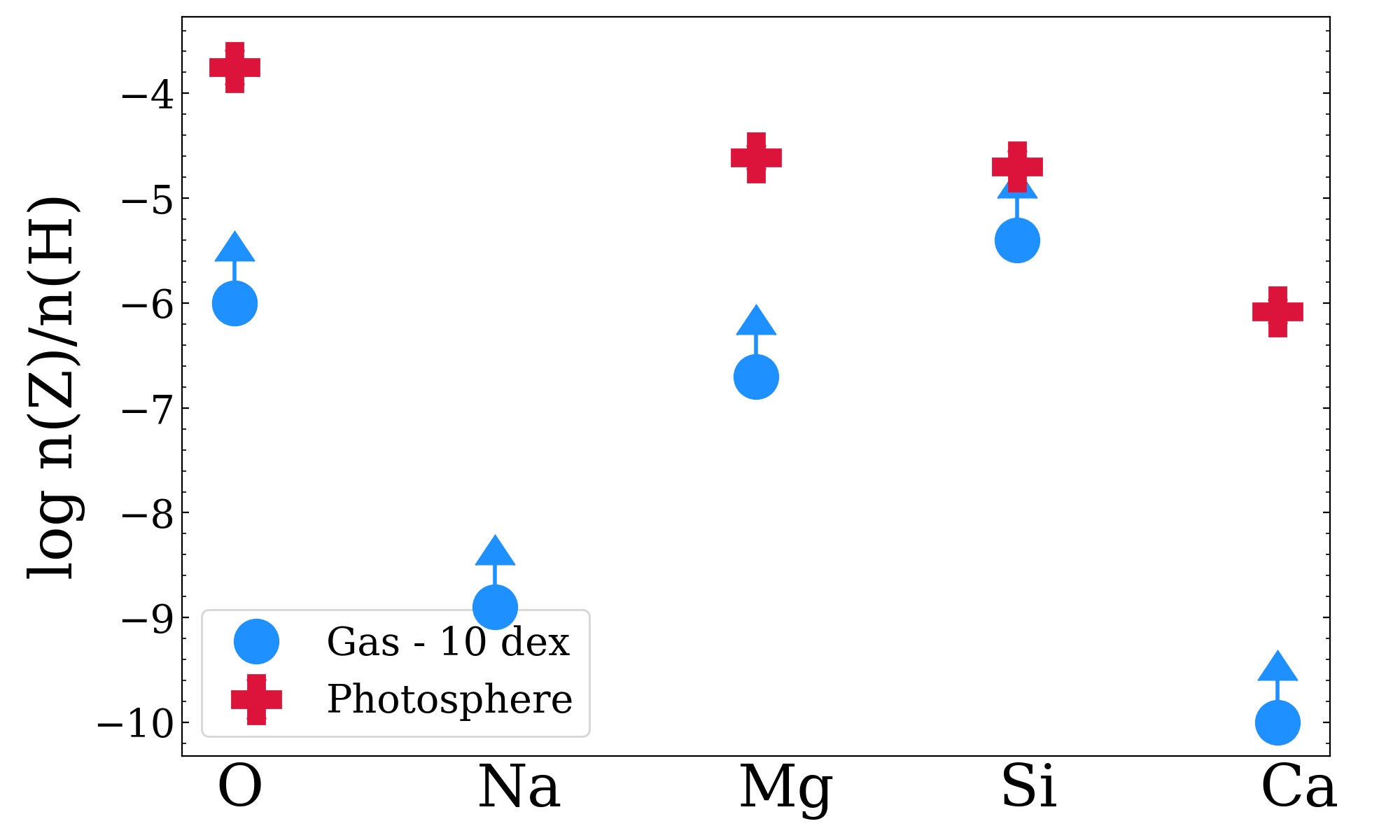}
    \caption{Logarithmic number ratios of different heavy elements with respect to {\color{black} hydrogen} in the circumstellar gas around Gaia~J0611$-$6931 (blue dots) {\color{black} and} the polluted atmosphere of Gaia~J0611$-$6931 (red stars). {\color{black} Relatively speaking, the upper limit to silicon is the most constraining in the circumstellar gas.} 
    }
\label{fig: composition}
\end{figure}

We can now compare the abundances of the circumstellar gas around Gaia~J0611$-$6931 with the photospheric abundances. The comparison is shown in Figure~\ref{fig: composition}. The photospheric abundance {\color{black} is taken from \citet{Rogers2024} without any additional correction}. The abundance difference between the build-up {\color{black}(the observed abundance in the white dwarf atmosphere equals that of the parent body)} and steady state {\color{black}(the observed abundance is the abundance of the parent body modified by differential settling, see details in \citealt{Dufour2016})} is within 0.2~dex, which does not make a difference in the interpretation here. 

Only upper limits were obtained for the elements in the circumstellar gas. Our \textsc{Cloudy} model assumes that the hydrogen abundance is the same as calcium, because no hydrogen emission line is detected. In Figure~\ref{fig: composition}, we arbitrarily subtracted 10 dex in the number ratios for circumstellar gas, to put the compositions of the photosphere and the gas in a similar scale.  {\color{black} For Gaia~J0611$-$6931, we found the upper limit to Si is the most constraining in the gas. 
}

\subsection{\color{black} Mass of the Gas Disk}

{\color{black}
The the mass of the gas disk can be estimated using the number densities in Table~\ref{tab:abundances}. The other assumption is that the disk extends from 20 to 100~R$_\mathrm{wd}$ with a constant scale height of 1~R$_\mathrm{wd}$. We can derive a lower limit on the mass of the gas disk mass to be 1.8 $\times$ 10$^{19}$~g, considering O, Si, Ca, Na and Mg. On the other hand, the total amount of heavy elements in the atmosphere of Gaia~J0611$-$6931 is 7.1 $\times$ 10$^{16}$, which is much smaller than the minimum mass of the gas disk. Taking the mass accretion rate of 6.57$\times$10$^8$ g s$^{-1}$ \citep{Rogers2024}, the minimum gas disk life time is 900~yr.

Our derived minimum gas disk mass of 1.8 $\times$ 10$^{19}$~g is close to the gas mass estimate of 10$^{21}$ g for another white dwarf SDSS J1228+1040 using emission line kinematics \citep{Goksu2023}. In other studies to model the circumstellar gas absorption around white dwarfs, the estimated gas mass is about 10$^{16}$ g for WD~1145+017 \citep{Fortin-Archambault2020} and 5 $\times$ 10$^8$ -- 1.3 $\times$ 10$^{16}$~g for WD 1124$-$293 \citep{Steele2021}. Compared to emission line detection, circumstellar absorption lines appear to be much more sensitive to small amount of circumstellar material around white dwarfs.
}

\subsection{Future Improvements}

In this paper, we developed a {\color{black} simple} model for the circumstellar gas around the white dwarf Gaia~J0611$-$6931 {\color{black} using \textsc{Cloudy}. The number densities of O, Si, Ca, Na and Mg are constrained and a minimum gas disk mass is also derived}. Here, we discuss the limitations of the current work and future improvements.

{\color{black} {\bf Modeling: disk geometry \& density profile \& scale height}. Our \textsc{Cloudy} model takes the simplest assumption that the gas disk is edge-on with a constant density profile and a constant scale height. This is a major uncertainty that affects the projected velocities, radiative transfer, disk temperature, and line flux. In addition, radiative transfer is only computed along the radial direction, and not in the vertical direction. Given the large optical depth in the radial direction, the line photons may escape from the disk surface. Future work is needed to explore these possibilities. }

{\bf Observation: emission lines with resolved profiles}. This study demonstrates the advantage of using infrared emission lines, which {\color{black} are more likely to be optically thin and therefore are better probes of the density of the gas}. However, due to the lower spectral resolution of the infrared data, the line profiles are not resolved and most of the infrared lines are heavily blended, limiting its usage. Future observations with a higher spectral resolution that resolves the profiles of the infrared emission lines will be particularly useful in constraining the parameters of the disk and the abundances. It will also be useful to search for additional emissions from other elements (such as those listed in Table~\ref{tab:predicted Lines}) that are currently not detected.

\section{Conclusions \label{sec: conclusions}}

In this paper, we present a model using the photoionization code \textsc{Cloudy} and an analytical method {\color{black} to study} the circumstellar gas around the white dwarf Gaia~J0611$-$6931. Emission lines from five different species -- O, Na, Mg, Si and Ca -- are detected around Gaia~J0611$-$6931. This detection is based on a comprehensive dataset that spans the ultraviolet, optical and infrared spectra. The emission lines are symmetric and stable, displaying no significant changes over a three year period and making it an ideal system for modeling. 

We provide here a first estimate for the composition of a refractory gaseous debris disk. Here is a summary of the main conclusions.

$\bullet$ The line flux for each emission line is measured and compared with the \textsc{Cloudy} output to constrain the abundances. For future work, we would like to encourage the use of the line flux (in favor of equivalent width), which is a more absolute way to characterize the emission line strength and allows disks to be directly compared to one another. 

$\bullet$ Our model {\color{black} takes some simple assumptions: (i) the gas is on a circular orbit, (ii) the disk is edge-on and radiative transfer is only computed along the radial direction, (iii) a constant density profile, (iv) a constant disk scale height}. Based on these assumptions, we find the best fit model requires the gas to extend from 20$R_\mathrm{wd}$ to 100$R_\mathrm{wd}$. The disk would be closer to the white dwarf if it is more inclined.

$\bullet$ {\color{black} We found most of the emission lines are completely saturated, and the line ratios approach the blackbody ratios. Only a lower limit to the number density can be derived. One complication is that the strength of a line from an element X depends on both the abundance of that particular element X and the other elements. Relatively speaking, silicon can be best constrained in the circumstellar gas with a lower limit of 10$^{10.3}$ cm$^{-3}$.
}

$\bullet$ The gas temperature is estimated to be {\color{black} 10,000~K to 4,500~K. Likely, the circumstellar gas is much closer to the white dwarf than the circumstellar dust disk}. We also placed a lower limit on the mass of the gas to be {\color{black} 1.8 $\times$ 10$^{19}$~g}. 

$\bullet$ Several improvements are needed in order to apply this method to a broader sample of white dwarfs with circumstellar gas. Modeling the composition of circumstellar gas {\color{black} may represent a novel approach} to constraining the composition of extrasolar planetesimals.

\vspace{5mm}
{\it Acknowledgements.} {\color{black} We would like to thank the referee Y. Wu, whose comments significantly improved the quality of the manuscript.} We would like to thank F. Lagos, O. Toloza, and B. G{\"a}nsicke on discussing the usage of \textsc{Cloudy}. We thank N. Jiang and B. Zuckerman for discussing the content of the paper.

This research is based on observations made with the NASA/ESA Hubble Space Telescope obtained from the Space Telescope Science Institute, which is operated by the Association of Universities for Research in Astronomy, Inc., under NASA contract NAS 5–26555. These observations are associated with program 16204.

S. Xu is supported by the international Gemini Observatory, a program of NSF’s NOIRLab, which is managed by the Association of Universities for Research in Astronomy (AURA) under a cooperative agreement with the National Science Foundation on behalf of the Gemini partnership of Argentina, Brazil, Canada, Chile, the Republic of Korea, and the United States of America. S. Blouin is a Banting Postdoctoral Fellow and a CITA National Fellow, supported by the Natural Sciences and Engineering Research Council of Canada (NSERC). L. Rogers acknowledges support of a Royal Society University Research Fellowship, URF\textbackslash R1\textbackslash 211421 and an ESA Co-Sponsored Research Agreement No. 4000138341/22/NL/GLC/my = Tracing the Geology of Exoplanets.

This research was supported in part by the National Science Foundation under Grant No. NSF PHY-1748958.

\end{CJK}

\vspace{5mm}

\software{Astropy \citep{Astropy2013,Astropy2018,Astroph2022}, Cloudy \citep{Ferland2017}, Matplotlib \citep{Matplotlib}}

\bibliography{WD.bib}

\end{document}